\documentclass[12pt]{article}
\usepackage{graphicx}
\usepackage{latexsym}
\usepackage{amssymb}
\newtheorem{theorem}{Proposition}
\begin{document}

%\parindent 0in
% The lines below are necessary for enumerating the equations
% according to the sections where they are.
\makeatletter
\@addtoreset{equation}{section}
\makeatother
\renewcommand{\theequation}{\thesection.\arabic{equation}}

\title{
Acoustic horizons in axially symmetric relativistic accretion 
 }
 \author{
 Hrvoje Abraham$^1$,
Neven Bili\'c$^1$
and
Tapas K.~Das$^2$
 \\
 $^1$Rudjer Bo\v{s}kovi\'{c} Institute, 10002 Zagreb, Croatia \\
$^2$ Harish Chandra Research Institute, 
 Allahabadh-211 019, India.\\
E-mail: ahrvoje@thphys.irb.hr, bilic@thphys.irb.hr, tapas@mri.ernet.in
 }

\maketitle

\begin{abstract}
Transonic accretion onto astrophysical objects
 is a unique example of
 analogue black hole realized in nature.
 In the framework of acoustic geometry
we study   axially symmetric accretion and wind
of a rotating astrophysical black hole
or of a neutron star
assuming 
isentropic flow of a fluid described by a polytropic
equation of state.
In particular we analyze the causal structure of  
multitransonic configurations with two sonic points and a shock.
Retarded and advanced  null curves clearly demonstrate
the presence of the acoustic black hole at regular sonic points and 
of the white hole at the shock.
We calculate the analogue surface gravity and the Hawking temperature for the
inner and  outer acoustic horizons.

\end{abstract}

\section{Introduction}
Analogue gravity has gained a lot of popularity in the last few years
\cite{nov}
(for a recent review and a comprehensive list of references, see \cite{barc})
for two obvious reasons:
First, it provides a laboratory for testing
some exotic effects of general relativity, such as
black holes, wormholes, warped space-time, etc.\ Second, since the scale of 
analogue gravity is not fixed by the 
Planck length,
there is hope that
quantum gravity effects, which normally are important
beyond the Planck scale only, may be  seen on larger length scales.

A particular example of analogue gravity is
acoustic geometry in which  an inhomogeneous fluid flow
provides a curved space-time background for the propagation
of sound waves. 
The most appealing aspect of acoustic geometry is
a relatively easy way to construct an acoustic horizon which is
an analogue of the  black hole horizon in general relativity.
In this way a number of interesting features which normally
are extremely difficult to observe in astrophysical
black holes, can be studied using an acoustic
analogue  of the black hole built in the laboratory.

In all  papers dealing with acoustic geometry
the flows with desired properties are artificially constructed.
So far, these constructions have been 
 only hypothetical since the 
necessary conditions for the actual observation of the horizon and the
measurements of the corresponding analogue Hawking temperature
are still difficult to realize in the laboratory.

Transonic accretion onto astrophysical black
holes is a unique example of
 analogue gravity realized in nature
  \cite{dascqg,das}.
 Spherical accretion  was
 first studied
  in terms of relativistic acoustic metric
  by Moncrief \cite{mon}.
  Recently, several aspects of analogue horizon radiation have been 
  investigated  
for a spherically symmetric general relativistic \cite{dascqg} and post-Newtonian
\cite{dasgrg} accretion flow.
In this paper  we study acoustic geometry of an axially symmetric accretion disk
round a rotating astrophysical black hole
or round a neutron star.
The flow considered in this paper is a relativistic generalization of the acoustic 
 black hole analogues.
To study acoustic geometry in most simple terms, we
assume 
isentropic accretion of the fluid described by a polytropic
equation of state.
Compared with a spherically symmetric flow, an axially symmetric flow
exhibits a more complex structure, 
which differs from the spherically symmetric flow in two aspects. 
First, in axially symmetric 
accretion 
the acoustic horizon and the stationary limit surface 
generally do not
coincide and second,
multitransonic configurations with more than one
acoustic horizons are possible.

In section  \ref{preliminaries}
 we outline the basic relativistic  
hydrodynamics necessary for the study of accretion
in terms of acoustic geometry.
In section \ref{accretion} we describe the 
relativistic equations governing the accretion flow.
In section \ref{acoustic} we discuss the formation of
a standing shock and acoustic horizons at sonic points
and in section \ref{causal} we study the acoustic causal structure
of multitransonic accretion and wind.
Section \ref{surface} is devoted to a calculation of the analogue surface gravity
and in section \ref{superradiance} we briefly discuss the acoustic superradiance.
Finally,
we conclude the paper with section \ref{conclude}.

\section{Preliminaries}
\label{preliminaries}
For the sake of completeness and consistency of the paper,
as a starting point we review some well-known basic results in relativistic
hydrodynamics and acoustics.
We  restrict attention to
a stationary gravitating isentropic fluid 
in a stationary background metric $g_{\mu\nu}$.
We work in units
$c=\hbar=G_{\rm N}=k_{\rm B}=1$
and use  the 
convention of positive-time-negative-space 
metric signature.

\subsection{Relativistic kinematics of the flow}

Let $u^{\mu}$ denote the 4-velocity field, i.e.,
the unit vector field tangent to the flow streamlines.
In general, a 4-velocity can be expressed
in terms of the 3-velocity components \cite{lan0,bil}
\begin{equation}
u^{\mu}
=
\gamma \left(
 \frac{1}{\sqrt{g_{00}}}
-\frac{g_{0j}v^j}{g_{00}};v^i
\right)  ,
\label{eq031}
\end{equation}
\begin{equation}
u_{\mu}=
 \gamma \left(
 \sqrt{g_{00}};
\frac{g_{0i}}{\sqrt{g_{00}}}-\gamma_{ij}v^j
\right)  ,
\label{eq032}
\end{equation}
with the induced
three-dimensional spatial metric
\begin{equation}
\gamma_{ij}=
\frac{g_{0i}g_{0j}}{g_{00}}-g_{ij}   \, ,
 \;\;\;\;\;\;\;
 i,j=1,2,3,
\label{eq033}
\end{equation}
and the Lorentz factor
\begin{equation}
\gamma^2= \frac{1}{1-v^2}, \;\;\;\;\; v^2=\gamma_{ij}v^i v^j;
\;\;\;\;\;\;
 i,j=1,2,3.
\label{eq2}
\end{equation}

Let $\{\Sigma_v\}$
denote a set of hypersurfaces defined by
\begin{equation}
v^2-a^2=0,
\label{eq135}
\end{equation}
where $a$ is a constant, $0\leq a <c$.
Each  $\Sigma_v$ is timelike since its normal
\begin{equation}
 n_{\mu}\propto
\partial_{\mu}v^2
\label{eq335}
\end{equation}
 is spacelike and 
may be normalized as $n^{\mu}n_{\mu}=-1$. 
The 4-velocity at each point
may be  decomposed into normal and tangential
components with respect to  $\Sigma_v$:
\begin{equation}
u^{\mu}=
v_{\perp}
\gamma_{\perp} n^{\mu}+
\gamma_{\perp} L^{\mu} ,
\label{eq037}
\end{equation}
where
\begin{equation}
L^{\mu}=\frac{1}{
\gamma_{\perp}}
(g^{\mu\nu}+n^{\mu}n^{\nu})
u_{\nu}\, ;
\;\;\;\;\;
\gamma_{\perp} = \frac{1}{\sqrt{1-v_{\perp}^2}} \, .
\label{eq237}
\end{equation}

Next, we consider a two-dimensional axisymmetric flow in a stationary
axisymmetric space-time. Physically, this may be a model for 
a black hole accretion disk, or for a draining flow in
an axially symmetric bathtub.
Using the notation $v^1\equiv v_r$, $v^2\equiv v_{\phi}$ and
$v^3\equiv v_z=0$,
 the 4-velocity  is given by
\begin{equation}
u^{\mu}
 =
\gamma \left(
 \frac{1}{\sqrt{g_{00}}}
-\frac{g_{0\phi}v_\phi}{g_{00}};v_r,v_{\phi},0
\right)  ,
\label{eq041}
\end{equation}
\begin{equation}
u_{\mu}=
 \gamma \left(
 \sqrt{g_{00}}; g_{rr}v_r,
\frac{g_{0\phi}}{\sqrt{g_{00}}}
-(\frac{g_{0\phi}^2}{g_{00}}-g_{\phi\phi})v_\phi,0
\right)  ,
\label{eq042}
\end{equation}
Then, the 3-velocity squared may be written as
\begin{equation}
v^2=v_{\parallel}^2-g_{rr}v_r^2,
\label{eq009}
\end{equation}
with
\begin{equation}
v_{\parallel}^2=(g_{0\phi}^2/g_{00}-g_{\phi\phi})v_{\phi}^2;
\;\;\;\;\;\;
\gamma_{\parallel}^2=\frac{1}{1-v_{\parallel}^2}\, .
\label{eq010}
\end{equation}
For a hypersurface of constant $v$ defined by (\ref{eq135})
the decomposition (\ref{eq037}) yields
\begin{equation}
v_{\perp}^2=-g_{rr}v_r^2\gamma_{\parallel}^2 \, ;
\;\;\;\;\;\;
\gamma^2=\gamma_{\perp}^2\gamma_{\parallel}^2.
\label{eq011}
\end{equation}

It is useful to define a {\em corotating frame} as a coordinate frame of reference
in which $v_{\parallel}$ (or $v_\phi$) vanishes and, as a consequence
$v_{\perp}=v$. 
 This frame is equivalent to
the frame comoving with $L^{\mu}$. In other words, in the corotating frame
the vector $L^{\mu}$ 
takes a simple form
\begin{equation}
L^{\mu}=
\frac{
\delta^{\mu}_{0}
}{\sqrt{g_{00}}}\,  ;
 \;\;\;\;\;\;\;
L_{\mu}=
\frac{
g_{\mu 0}
}{\sqrt{g_{00}}}\, .
\label{eq230}
\end{equation}
The transition to a corotating frame 
 is achieved by a simple 
coordinate transformation
\begin{equation}
%d\phi=d\phi'+\frac{u^\phi}{u^0} dt \, .
d\phi=d\phi'+\Omega dt \, ,
\label{eq231}
\end{equation}
where the angular velocity $\Omega$ is defined as
\begin{equation}
\Omega= \frac{u^{\phi}}{u^0}\, .
\label{eq001}
\end{equation}

More generally, we may define a 
{\em coorbiting frame}  as 
a coordinate frame of reference comoving with $L^{\mu}$.
Hence  an axially symmetric coorbiting frame
is corotating.

\subsection{Acoustic geometry}
\label{geometry}
The acoustic metric tensor
and its inverse are defined by
\cite{mon,bil}
\begin{equation}
G_{\mu\nu} =\frac{n}{wc_s}
[g_{\mu\nu}-(1-c_s^2)u_{\mu}u_{\nu}] ,
\label{eq022}
\;\;\;\;\;
G^{\mu\nu} =\frac{wc_s}{n}
\left[g^{\mu\nu}-(1-\frac{1}{c_s^2})u^{\mu}u^{\mu}
\right]\, ,
\label{eq029}
\end{equation}
where $n$ is the particle number, $w=(\rho+p)/n$ the specific enthalpy
of the fluid
and $c_s$ the speed of sound.
In analogy to the event horizon in general relativity,
the {\em acoustic horizon} may be defined as a
time-like  hypersurface 
whose normal $n_{\mu}$ is
null with respect to the acoustic metric,
i.e.,
\begin{equation}
G^{\mu\nu}n_{\mu}n_{\nu}=0.
\label{eq207}
\end{equation}
Equivalently, we may define the acoustic horizon as a hypersurface
defined by the equation \cite{bil}
\begin{equation}
v_{\perp}^2-c_s^2=0.
\label{eq208}
\end{equation}

The metric defined by (\ref{eq029}) is very general
and with no further assumptions the determination of acoustic horizons
may be  nontrivial. However, the matter may greatly simplify
if we assume a certain symmetry, such that the analogy with the 
familiar stationary axisymmetric black hole may be drawn.

We consider the acoustic geometry that
satisfies
       the following assumptions:
\begin{enumerate}
\item
The acoustic metric is stationary.
\item
The flow is symmetric under
the transformation that generates
a displacement on the hypersurface $\Sigma_v$ of constant $v$ along the projection of
the flow velocity.
\item
The metric $g_{\mu\nu}$ is invariant under the
above transformation.
\end{enumerate}
The last two statements are equivalent to saying that
a displacement
 along the projection of the flow velocity on $\Sigma_v$
is an isometry.
The axisymmetric flow considered in section
\ref{geometry} is an example that fulfils the above
requirements.

It may be easily shown that,
for an axially symmetric flow,
  the metric discriminant
\begin{equation}
{\mathcal D}\equiv G_{0\phi}^2-G_{00}G_{\phi\phi}
\label{eq209}
\end{equation}
vanishes at the acoustic horizon.

In analogy to the Kerr black hole we may also define the
{\em acoustic ergo region}
as a region in space-time where
the stationary Killing vector $\xi$
becomes space-like.
The magnitude  of $\xi^{\mu}$, defined with respect to the acoustic metric,
is given by
\begin{equation}
\parallel\! \xi \!\parallel =
G_{\mu\nu} \xi^\mu \xi^\nu =\frac{n}{wc_s}
 \xi^{\mu}\xi_{\mu}
\gamma^2 (c_s^2-v^2).
\label{eq034}
\end{equation}
This becomes negative when the magnitude of
the flow velocity $v$ exceeds the speed of sound $c_s$.
  The boundary of the ergo-region 
  defined by the equation
\begin{equation}
v^2-c_s^2=0.
\label{eq035}
\end{equation}
 is a hypersurface  called {\em stationary limit surface}
 \cite{haw73}.
 Equivalently, the stationary limit surface may
 be defined as a hypersurface  at which
 \begin{equation}
G_{00}=0.
\label{eq036}
\end{equation}
 
In the following sections we often use the terms
{\em subsonic}, {\em supersonic} and {\em transonic}.
These notions
in relativistic hydrodynamics
make sense
only with respect to a specified frame of reference.
For example, a flow which is supersonic with respect to an observer at rest
is subsonic with respect to 
a comoving observer.
In an axially symmetric flow
it is convenient to 
use the corotating frame of reference,
i.e., the frame in which $v_{\phi}$ vanishes, to define
a locally subsonic or supersonic flow in the following way:
the flow at a radial distance $r$ is said to be  subsonic or supersonic   
if  $v(r) < c_s(r)$ or 
$v(r) > c_s(r)$, respectively.
It may be shown that the above definition is
consistent with another
definition  valid in
any other reference frame that preserves 
temporal and axial isometry:
the axisymmetric flow is said to be
locally subsonic or supersonic if
$v_{\perp}(r) < c_s(r)$ or 
$v_{\perp}(r) > c_s(r)$, respectively.
Alternatively, as we  demonstrate in section \ref{causal},
we may use equation (\ref{eq209})
to define a flow with ${\cal D} > 0$  as subsonic
and with ${\cal D} < 0$ as  supersonic. 
Finally,  a flow is said to be {\em transonic}
 if a transition from a subsonic to a supersonic flow region,  
 or vice versa, takes place. 
 The point of transition is called the {\em sonic point}
 if the transition is continuous.
 A discontinuous supersonic to subsonic transition is called {\em shock}. 
 
 \subsection{Propagation of acoustic perturbations}
 \label{propagation}
Propagation of acoustic waves in an inhomogeneous 
fluid moving in curved space-time is governed
by the massless wave equation
\begin{equation}
\frac{1}{\sqrt{-G}}\,
\partial_{\mu}
(\sqrt{-G}\,
G^{\mu\nu})
\partial_{\nu}\varphi=0,
\label{eq028}
\end{equation}
where 
\begin{equation}
G =
\det G_{\mu\nu} \, .
\label{eq025}
\end{equation}
It may be shown that
equation (\ref{eq028}) follows from the free wave equation
that describes the propagation of acoustic perturbations in
a homogeneous fluid in flat space-time
\begin{equation}
\eta_{\mu\nu}
\frac{\partial}{\partial x^\mu}
\frac{\partial}{\partial x^\nu}\varphi
=0,
\label{eq129}
\end{equation}
with $x^0=c_st$, $x^i=(x,y,z)$
and $\eta_{\mu\nu}=(1,-1,-1,-1)$.
Equation (\ref{eq028}) is obtained
by replacing $\eta_{\mu\nu}$ by $G^{\mu\nu}$ 
and $\partial/\partial x^{\mu}$ by the covariant derivative $\nabla_\mu$ 
associated with the metric $G_{\mu\nu}$.
In this way we introduce the  ``acoustic equivalence principle"
in full analogy with general relativity. 

We  also define the stress tensor of the field $\varphi$ 
\begin{equation}
T_{\mu\nu}=
\partial_\mu\varphi\,
\partial_\nu\varphi
-\frac{1}{2} 
G^{\alpha\beta}
\partial_\alpha\varphi\,
\partial_\beta\varphi\,
G_{\mu\nu}
\label{eq329}
\end{equation}
which satisfies 
$G^{\mu\alpha}\nabla_\alpha T_{\mu\nu}=0$.

By standard arguments \cite{lan0,wal} it follows that the
lines of  acoustic propagation
are null geodesics with respect to
the acoustic metric.
 In other words, the wave vector
$k^{\mu}$, which is timelike and may be normalized to unity,
is  null with respect to the acoustic metric, i.e.,
\begin{equation}
 G_{\mu\nu} k^\mu k^\nu=0, 
\label{eq030}
\end{equation}
and satisfies the acoustic geodesic equation
\begin{equation}
 k^\mu \nabla_{\mu} k^\nu=0, 
\label{eq131}
\end{equation}
where
\begin{equation}
 \nabla_\mu k^\nu={k^\nu}_{;\mu}+\Gamma^\nu_{\mu\alpha}k^\alpha 
\label{eq132}
\end{equation}
and
\begin{equation}
 \Gamma^\nu_{\mu\alpha}=\frac{1}{2} G^{\nu\beta}
 (G_{\mu\beta;\alpha}+G_{\alpha\beta;\mu} -G_{\mu\alpha;\beta}).
\label{eq133}
\end{equation}
However, as we have said, $k^{\mu}$ is timelike 
with respect to the space-time metric, so
acoustic perturbations propagate along timelike curves.

Next we show that
the perturbations that propagate parallel to the 
streamlines 
cross the horizon and those that propagate in the opposite direction are 
trapped. 
We assume that the space-time metric is stationary and the flow is steady.  
At any fixed 
time $t_0$  the flow congruence crosses  the
two-dimensional 
surfaces \{$v^2$=const, $t=t_0$\}.
In particular, the acoustic horizon 
defined by $v_{\perp}^2=c_s^2$ at fixed $t=t_0$
is one of these
surfaces.
The hypersurface  $v_{\perp}^2=c_s^2$  is of course
timelike since its normal $ \partial_{\mu} (v_\perp^2-c_s^2)$ is a 
spacelike vector.
Let $k^{\mu}$ be  a timelike unit vector tangent to the lines of acoustic 
propagation.
Then $k^{\mu}$ may be decomposed in a similar way as $u^\mu$ in equation 
(\ref{eq037}),
\begin{equation}
k^{\mu}=
k\gamma_k
 n^{\mu}+
 \gamma_k
 l^{\mu} ,
\label{eq137}
\end{equation}
where 
\begin{equation}
l^{\mu}=\frac{1}{
\gamma_{k}}
(g^{\mu\nu}+n^{\mu}n^{\nu})
k_{\nu}\, ;
\;\;\;\;\;
\gamma_{k} = \frac{1}{\sqrt{1-k^2}} \, .
\label{eq337}
\end{equation}
Now we restrict attention to the perturbations
for which $l^\mu=L^\mu$. 
In the coorbiting frame
defined in section \ref{geometry}, 
these perturbations propagate either parallel
or antiparallel to the streamlines,
i.e.,
the spatial part of $k^\mu$ is parallel or antiparallel to 
the 3-velocity of the fluid.
 As a consequence of (\ref{eq030}) we have
$k^{\mu}u_{\mu}=\gamma_{c_s}$. 
This equation states the fact that 
acoustic perturbations propagate with the velocity of sound $c_s$
with respect to the fluid.
At the horizon, as may be easily shown, 
$k=2c_s/(1 + c_s^2)$ for the perturbations going parallel,
and $k=0$  for those going antiparallel to the streamlines.
Since in the first case we have  $k^\mu n_\mu \neq 0$,
the acoustic perturbations that propagate parallel to
the streamlines
will cross the horizon.
The perturbations going antiparallel
 will satisfy  $k^\mu n_\mu = 0$ at the horizon
and hence will be trapped.

\subsection{Quantization of phonons and the Hawking effect}
\label{quantization}
The purpose of this section is to demonstrate how the
quantization of phonons in the  presence of
the acoustic horizon yields
acoustic Hawking radiation.
The acoustic perturbations considered here are classical
sound waves or {\em phonons} that
 satisfy the massles wave 
equation (\ref{eq028})
in curved background  with the metric $G_{\mu\nu}$
given by (\ref{eq029}).
Irrespective of the underlying microscopic structure
acoustic perturbations are quantized.
A precise quantization scheme for an analogue gravity
system may be rather involved \cite{unr}.
However, at the scales larger than
the atomic scales below which a perfect fluid description breaks down,
the atomic substructure may be neglected and
the field may be considered elementary. Hence,
the quantization proceeds in the same way as in the case of a
scalar field in curved space \cite{bir}
with a suitable UV cutoff for the scales below a typical
atomic size of a few \AA. 

For our purpose, the most convenient
quantization prescription is the Euclidean path integral
formulation.
Consider a 2+1-dimensional disc geometry.
The equation of motion (\ref{eq028})
follows from the  variational principle applied to
the action functional 
\begin{equation}
S[\varphi]=\int dtdrd\phi\, 
\sqrt{-G}\,
G^{\mu\nu}
\partial_{\mu}\varphi
\partial_{\nu}\varphi\, .
\label{eq228}
\end{equation}
 We define the functional integral 
\begin{equation}
Z= \int {\cal D}\varphi e^{-S_{\rm E}[\varphi]} ,
\label{eq229}
\end{equation}
where $S_{\rm E}$ is the Euclidean  action 
obtained from (\ref{eq228}) by setting
%\begin{equation}
$t=i\tau$
%\label{eq232}
%\end{equation}
and continuing the Euclidean time $\tau$ from imaginary to real values.
For a field theory at zero temperature, the integral
over $\tau$ extends up to infinity.
Here,
 owing to the presence of the acoustic horizon,
 the integral over $\tau$ 
 will be cut at the inverse Hawking temperature $2\pi/\kappa$
 where $\kappa$ denotes the analogue surface gravity.
 To illustrate how this happens, consider, for simplicity, a nonrotating 
 fluid ($v_\phi=0$) in an arbitrary static, spherically symmetric space-time.
 It may be easily shown that the acoustic metric takes the form
 \begin{equation}
ds^2=\frac{c_s^2-v^2}{1-v^2}g_{00}\, dt^2 -2 v\frac{1- c_s^2}{1-v^2}\sqrt{-g_{00}g_{rr}}\,
drdt +\frac{2-c_s^2v^2}{1-v^2}g_{rr}\, dr^2 -r^2 d\phi^2\, ,
\label{eq232}
\end{equation}
where the background metric coefficients $g_{\mu\mu}$ are functions 
of $r$ only, and we have omited the irrelevant conformal factor
$n/(w c_s)$.
Using the coordinate transformation
\begin{equation}
dt\rightarrow dt+\frac{v(1- c_s^2)}{c_s^2-v^2}
\sqrt{-\frac{g_{rr}}{g_{00}}}\, dr
\label{eq233}
\end{equation}
we remove the off-diagonal part from (\ref{eq232}) and obtain
\begin{equation}
ds^2=\frac{c_s^2-v^2}{1-v^2}g_{00}\, dt^2 +
\frac{1}{1-v^2}\left[2-c_s^2v^2+
\frac{v^2(1- c_s^2)^2}{(c_s^2-v^2)}\right] g_{rr}\, dr^2 
 -r^2 d\phi^2.
\label{eq234}
\end{equation}
Next, we evaluate the metric near the acoustic horizon at
$r=r_{\rm h}$ using the expansion in $r-r_{\rm h}$ at first order
 \begin{equation}
c_s^2-v^2\approx 2 c_s \left. \frac{\partial}{\partial r}(c_s-v)
\right|_{\rm h}(r-r_{\rm h})
\label{eq235}
\end{equation}
and making the substitution
\begin{equation}
r-r_{\rm h}=-\frac{1}{2c_s (1-c_s^2)}
\frac{1}{g_{rr}}
\left. \frac{\partial}{\partial r}(c_s-v)\right|_{\rm h} \rho^2,
\label{eq236}
\end{equation}
where $\rho$ denotes a new radial variable.
Neglecting the subdominant terms  in the square brackets in (\ref{eq234})
and setting $t=i\tau$, we obtain the Euclidean metric in the form
\begin{equation}
ds_{\rm E}^2=\kappa^2 \rho^2 d\tau^2 +d\rho^2 +r_{\rm h}^2
 d\phi^2\, ,
\label{eq437}
\end{equation}
where 
\begin{equation}
\kappa=\frac{1}{1-c_s^2}
\sqrt{-\frac{g_{00}}{g_{rr}}}
\left. \frac{\partial}{\partial r}(c_s-v)\right|_{\rm h}\, .
\label{eq238}
\end{equation}
Hence, the metric near $r=r_{\rm h}$ is the product of the metric on S$^1$
and the Euclidean Rindler space-time 
\begin{equation}
ds_{\rm E}^2=d\rho^2 + \rho^2 d(\kappa \tau)^2 .
\label{eq239}
\end{equation}
With the periodic identification 
$\tau\equiv \tau+2\pi/\kappa$, the metric (\ref{eq239})
describes $\mathbb R^2$ in plane polar coordinates. 

Furthermore, making the substitutions
$\rho=e^{\kappa x}/\kappa$ and $\phi=y/r_{\rm h}+\pi$,
the Euclidean action takes the form of the 
2+1-dimensional free scalar field action
at nonzero temperature
\begin{equation}
S_{\rm E}[\varphi]=\int_0^{2\pi/\kappa} d\tau 
\int_{-\infty}^{\infty}dx
\int_{-\infty}^{\infty} dy \frac{1}{2} (\partial_{\mu} \varphi)^2,
 \label{eq240}
\end{equation} 
where we have set
the upper and lower bounds of the integral over $dy$
to $+\infty$ and $-\infty$, respectively,
assuming  that $r_{\rm h}$ is sufficiently large.
Hence, the functional integral $Z$ in (\ref{eq229}) 
is evaluated over the fields $\varphi(x,y,\tau)$ that are periodic in
$\tau$ with period  $2\pi/\kappa$.
In this way, the functional $Z$ is just the
 partition function for a grandcanonical ensemble of free 
 massless bosons
  at the 
Hawking temperature
$T_{\rm H}=\kappa/(2\pi)$.
However, the radiation spectrum will not be exactly thermal
since we have to cut off the scales below the atomic scale
\cite{unr2}. The choice of the cutoff and the deviation of 
the acoustic radiation spectrum from the thermal spectrum is
closely related to the so-called {\em transplanckian problem}
of Hawking radiation \cite{jac}.

 \subsection{Conservation laws}
Our basic equations are the continuity equation
\begin{equation}
(nu^{\mu})_{;\mu}=
\frac{1}{\sqrt{-g}}
\partial_{\mu}({\sqrt{-g}}\,  n u^{\mu})=0 
\label{eq101}
\end{equation}
and the energy-momentum
conservation
\begin{equation}
{T^{\mu\nu}}_{;\nu}=0.
\label{eq102}
\end{equation}
  Applied to  a perfect, isentropic fluid, this equation yields
 the relativistic Euler equation \cite{lan2}
 \begin{equation}
 u^{\nu}(wu_{\mu})_{;\nu}
-\partial_{\mu}w=0.
\label{eq103}
\end{equation}

The quantities $c_s$, $u_{\mu}$ and $T_{\mu\nu}$ may be discontinuous at
a hypersurface $\Sigma$. In this case, equations (\ref{eq101}) and (\ref{eq102})
imply
\begin{equation}
\left[\left[nu^{\mu}\right]\right] n_{\mu}=0,
\label{eq104}
\end{equation}
\begin{equation}
\left[\left[T^{\mu\nu}\right]\right]n_{\nu}=0,
\label{eq105}
\end{equation}
where $[[f]]$ denotes the discontinuity of $f$ across $\Sigma$, i.e.,
\begin{equation}
\left[\left[f\right]\right]=f_2 -f_1,
\label{eq106}
\end{equation}
with $f_2$ and $f_1$ being the boundary values on the two sides of
$\Sigma$. For a perfect fluid, equations  (\ref{eq104}) and (\ref{eq105})
may be written as
\begin{equation}
\left[\left[n v_{\perp}\gamma_{\perp}\right]\right]=0,
\label{eq401}
\end{equation}
\begin{equation}
\left[\left[T_{0\mu}n^{\mu}\right]\right]=
\left[\left[(p+\rho)u_0 v_{\perp}\gamma_{\perp} \right]\right]=0,
\label{eq402}
\end{equation}
\begin{equation}
\left[\left[T_{\mu\nu}n^{\mu}n^{\nu}\right]\right]=
\left[\left[(p+\rho)v_{\perp}^2\gamma_{\perp}^2+p \right]\right]=0.
\label{eq403}
\end{equation}
Equations (\ref{eq401})-(\ref{eq403}) are referred to as
the relativistic Rankine-Hugoniot equations \cite{tau}.

\subsection{Constants of motion}
The acoustic metric and the background space-time metric
generally do not share the isometries. For example,
the axial symmetry of
  a tilted 
accretion disc round a rotating black hole 
obviously does not coincide with that of the black hole.

Here  we assume that the flow
 and 
the space time metric $g_{\mu\nu}$ share a specific symmetry generated by a
Killing vector $\chi^\mu$.
\begin{theorem}
 Let $\chi^\mu$ be a Killing vector field and let the flow be
 isentropic and invariant under the group of transformations
  generated 
 by $\chi^\mu$.
 Then the quantity
$w\chi^{\mu}u_{\mu}$
%$w=(p+\rho)/n$ being the specific enthalpy,
 is constant along the stream line.
\end{theorem}
{\bf Proof}:
Multiplying
the  Euler equation
(\ref{eq103})
by $\chi^{\mu}$ we obtain
\begin{equation}
 u^{\nu}\partial_{\nu}(w\chi^{\mu}u_{\mu})
-u^{\mu}u^{\nu}w\chi_{\mu ;\nu}-\chi^{\mu}\partial_{\mu}w =0.
\label{eq006}
\end{equation}
The second term in (\ref{eq006})
 vanishes by Killing's equation and the last term
vanishes by symmetry. Hence
\begin{equation}
 u^{\nu}\partial_{\nu}(w\chi^{\mu}u_{\mu})
 =0,
\label{eq007}
\end{equation}
as was to be proved. {\LARGE $\Box$}

\begin{itemize}
\item[]
{\bf Remark}  
Since,  by assumption, the flow shares the symmetry
of $g_{\mu\nu}$,
the Killing vector $\chi^{\mu}$
is also Killing with respect to the acoustic
metric.
\end{itemize}
\subsection{Polytropic gas}
Consider a gas of particles of mass $m$ 
described by the polytropic equation of state
\begin{equation}
p=m {\cal K} n^{\Gamma},
\label{eq216}
\end{equation}
where ${\cal K}$ is a constant which, as we shall shortly see, 
is related to the
specific entropy of the gas.
Using this equation and
the thermodynamic identity
\begin{equation}
dw=Td\left(\frac{s}{n}\right)+\frac{1}{n} dp,
\label{eq316}
\end{equation}
the energy density $\rho$ and the entropy density $s$ may also be expressed
in terms of $n$.
For an isentropic flow, it follows
\begin{equation}
\rho=m n\left(1+\frac{\cal K}{\Gamma-1} n^{\Gamma-1}\right).
\label{eq317}
\end{equation}

For an adiabatic process, the entropy density is proportional to the
particle number density, i.e., $s=\sigma n$. Without further assumptions, the  entropy density per particle 
$\sigma$ is an undetermined constant. 
If, in addition to (\ref{eq216}),
the Clapeyron equation 
for an ideal gas holds
\begin{equation}
p=nT  ,
\label{eq320}
\end{equation}
and assuming that
$\gamma$ is temperature independent,
then 
the entropy per particle is given by
\cite{lan1}
\begin{equation}
\sigma=\frac{1}{\Gamma -1} \log(m{\cal K})+\zeta +\frac{\Gamma}{\Gamma-1}\, ,
\label{eq322}
\end{equation}
where $\zeta$ is a constant that depends only on the chemical 
composition of the gas.
Hence, the quantity ${\cal K}$ measures the specific entropy.

 It is convenient to express all relevant thermodynamic quantities in
 terms of the adiabatic  speed of sound defined by
 \begin{equation}
c_s^2=\frac{n}{w}\left. \frac{\partial w}{\partial n}\right|_{\sigma} \, .
\label{eq213}
\end{equation}
We find
\begin{equation}
\frac{w}{m}=\frac{\Gamma-1}{\Gamma-1-c_s^2},
\label{eq217}
\end{equation}
\begin{equation}
n=\eta {\cal K}^{-\frac{1}{\Gamma-1}} \left(\frac{\Gamma-1}{c_s^2}-1\right)^{-\frac{1}{\Gamma-1}},
\label{eq318}
\end{equation}
\begin{equation}
p=m\eta^{\Gamma} {\cal K}^{-\frac{1}{\Gamma-1}}
\left(\frac{\Gamma-1}{c_s^2}-1\right)^{-\frac{\Gamma}{\Gamma-1}},
\label{eq319}
\end{equation}
where
\begin{equation}
\eta=\left(\frac{\Gamma-1}{\Gamma}\right)^{\frac{1}{\Gamma-1}}.
\label{eq420}
\end{equation}
%
%\begin{appendix}
\section{Accretion disk}
\label{accretion}
For a flow of matter with a non-zero angular momentum density, spherical
symmetry is broken and accretion phenomena are studied employing
an axisymmetric configuration.
Accreting matter is thrown into circular orbits,
leading to the formation of  {\it accretion disks} around 
astrophysical black holes or neutron stars. 
The pioneering contribution to study the properties of general
relativistic accretion disks is attributed to two classic papers by Bardeen, Press and
Teukolsky \cite{bar} and Novikov and Thorne \cite{nt73}.

If the accreting material
is assumed to be at rest far from the accreting black hole, the flow must exhibit
 transonic behaviour since the velocity of fluid particles necessarily
approaches the speed of light as they approach the black-hole event horizon.
 For certain
 values of the intrinsic angular
momentum density of the accreting material, the number of sonic points,
 unlike in spherical
accretion, may exceed one, and accretion is called {\em multi-transonic}.
Study of
such multi-transonic flows was initiated by Abramowicz and Zurek \cite{zurek}.
Subsequently, multi-transonic phenomena in black hole accretion disks and formation  of shock waves as a consequence of these phenomena, 
have been studied by a 
number of authors (for further details and for  references
see, e.g., \cite{das,fukue,apj1,dasmnras}).  

\subsection{Axisymmetric disc accretion}
\label{axisymmetric}
In this section, we outline the formalism for calculating
the location of  sonic points  as a function
of the specific flow energy ${\cal E}$,
the specific
angular momentum $\lambda$ and the polytropic index $\Gamma$
\cite{apj1,dasmnras}
applied to
  axisymmetric accretion.
We assume that the accretion flow is described by
 a disk placed in the equatorial plane
 of a rotating astrophysical object,
e.g., a Kerr black hole
or a neutron star, of mass $M$.
Specifying the metric to be stationary and axially symmetric,
 the two
generators
 $\xi^{\mu}$ and $\phi^{\mu}$ of the temporal and
axial isometry, respectively,  satisfy  equation (\ref{eq007}) and hence
the quantities ${\cal E}=wu_0$ and $L=wu_{\phi}$ are constant
along the streamlines.
Given the equation of state of the fluid, the constants of motion
$\cal E$ and $L$ fully determine the accretion flow.
It is convenient to introduce another constant
\begin{equation}
\lambda = -\frac{L}{\cal E}=-\frac{u_{\phi}}{u_0}\, ,
\end{equation}
which is called {\it specific angular momentum}.
It may be easily shown that
the angular velocity $\Omega$ defined 
by (\ref{eq001})
is expressed in terms of $\lambda$ as
\begin{equation}
\Omega= -\frac{g_{0\phi}+\lambda g_{00}}{g_{\phi\phi}+\lambda g_{0\phi}}.
\label{eq002}
\end{equation}
With the help of this equation,
either from  the normalization condition $u^{\mu}u_{\mu}=1$
or directly from (\ref{eq001}) with (\ref{eq010})-(\ref{eq011}),
we find
\begin{equation}
u_0^2= \gamma^2_{\perp}\frac{\Delta}{B}\, ,
\label{eq003}
\end{equation}
where, using the notation of \cite{bar,wal},
$\Delta$ and $B$ expressed 
in Boyer-Lindquist coordinates are 
\begin{equation}
\Delta=
g_{0\phi}^2-g_{00}g_{\phi\phi}=r^2-2Mr+a^2,
\label{eq206}
\end{equation}
\begin{equation}
B=-
(g_{\phi\phi}+2\lambda g_{0\phi}+ \lambda^2 g_{00})=
\left(r^{2}+a^{2}+\frac{2Ma^{2}}{r}\right) 
-4\lambda\frac{Ma}{r}
-\lambda^2 \left(1-\frac{2\,M}{r}\right).
\label{eq212}
\end{equation}

It is useful to introduce the particle accretion rate $\dot{N}$
which,
owing to the particle number conservation, does not depend on the radial coordinate $r$.
Besides assuming axial symmetry, we also assume that the
disk thickness $h$ is constant and the
distribution of matter in the disk does not depend on
the $z$-coordinate along the rotation 
axis\footnote{A  physically more realistic 
calculation would involve 
 solving the relativistic Euler equation 
 (\ref{eq103})
in the vertical direction 
\cite{dasmnras}.
This work
 is in progress and will be presented elsewhere.}.
 We start from the expression
\begin{equation}
N=\int_{\Sigma} n u^{\mu} d\Sigma_{\mu}\,
\label{eq201}
\end{equation}
for the number of particles the worldlines of which cross
a hypersurface $\Sigma$. Defining the hypersurface at rest
as a cylinder of radius $r$ and height $h$ and
by making use of the axial symmetry
and the continuity equation (\ref{eq101})
integrated over the disk volume,
we find
\begin{equation}
\dot{N}/h=-2\pi \sqrt{-g} n u^r \,.
\label{eq204}
\end{equation}
Using (\ref{eq011})  we obtain
\begin{equation}
\dot{N}/h=-2\pi \sqrt{\Delta} n v_{\perp} \gamma_{\perp} \, .
\label{eq205}
\end{equation}

Thus, the  two conditions
\begin{equation}
\dot{N}/h=-2\pi n v_{\perp} \gamma_{\perp}\Delta^{1/2}={\rm const} \,,
\label{eq210}
\end{equation}
\begin{equation}
{\cal E}\equiv wu_0=w \gamma_{\perp} \frac{\Delta^{1/2}}{B^{1/2}}
={\rm const} \,,
\label{eq211}
\end{equation}
together with the equation of state, fully determine 
the kinematics of the accretion disk. In other words, 
given the equation of state, 
e.g., in the form of 
(\ref{eq216}),
the metric and a
set of parameters \{$\lambda,{\cal E},\dot{N}/h$\},
the flow variables $v_{\perp}$ and $c_s$ may be
determined as functions of $r$.

\subsection{Non-axisymmetric accretion}
In this work, we concentrate only on axisymmetric fluid accretion, 
for which the orbital angular momentum of the entire disc plane 
remains aligned with the angular momentum of the compact object
at the origin.
 In a strongly coupled binary system (with 
a compact object as one of the components), accretion may experience 
a non-axisymmetric potential because the secondary donor star 
exerts a tidal force on the accretion disc round the 
compact primary. In general, a non-axisymmetric tilted disc may form if
the accretion takes place out of the symmetry plane of the spinning 
compact object. The matter in such a misaligned disc will experience a 
torque due to the general relativistic Lense-Thirring effect 
\cite{1}, leading to the precession of the inner disc
plane. The differential precession may cause stress and 
dissipative effects in the disc. If the torque remains strong enough 
compared with the internal viscous force, the inner region of the initially
tilted disc may be forced to realign itself with the   
symmetry plane of the central accretor. This phenomenon of
partial realignment, out to a certain radial distance known as the 
{\em transition radius} or the {\em alignment radius}, of the initially 
non-axisymmetric disc is known as the ``Bardeen-Petterson effect''
\cite{2}. The transition radius can be obtained
by balancing the precession and the inward drift or the viscous time
scale. 

An astrophysical accretion disc subject to the Bardeen-Petterson effect 
becomes {\em twisted} or {\em warped}. A large scale twist in the disc 
modifies the emergent spectrum and can influence the direction of the 
quasar and micro-quasar jets emanating out from the inner region of 
the accretion disc \cite{3}. 

A twisted disc may be thought of 
 as an ensemble of rings of increasing
radii, for which the variation of the direction of the orbital angular
momentum occurs smoothly while crossing the alignment radius. A system
of equations describing such a twisted disc has been formulated \cite{5},
 and the time scale required for a Kerr black hole
to align its  angular momentum with that of the initially 
misaligned accretion disc, has  been estimated \cite{8}.
 Numerical simulations using the three-dimensional Newtonian 
 smoothed particle hydrodynamics (SPH)
 code \cite{9}
 as well as a fully
relativistic framework \cite{10} reveal the geometric 
structure of twisted discs. 

Note, however, that as long as the acoustic horizon forms at a radial 
length scale smaller than that of the alignment radius, typically
of order
100 - 1000 Schwarzschild radii according to the original estimate of
Bardeen and Petterson \cite{2}, one need not implement the non-axisymmetric
geometry to study the analogue effects.

\section{Acoustic horizon}
\label{acoustic}
The flow considered here may be regarded as a relativistic generalization
of the two-di\-men\-sional draining bathtub flow with a sink at the origin.
Instead of a sink, at the origin we have  an astrophysical black hole 
or a neutron star playing the role of the sink.
Hence, our accretion flow is a general-relativistic black-hole analogue.

In the axially symmetric flow, the
acoustic horizon is a stationary circle of radius
$r_{\rm h}$ where, according to (\ref{eq208}),  
the radial component of the flow experiences a transition from
a subsonic to a supersonic region.
As we shall shortly demonstrate,
in our perfect polytropic gas model
such a transition may be realized in three
ways:
\begin{description}
\item[Smooth transition] Both $c_s$ and $v_{\perp}$ and their derivatives are
 regular at the transition point. In this case, the transition point is called 
{\em regular sonic point} or simply {\em sonic point}.
 The points  known in the literature as {\em saddle}-type or
X-type critical points
belong to this class.
The isolated sonic points known as 
{\em centre}-type or O-type also belong to this class. 
\item[Singular transition]
Both $c_s$ and $v_{\perp}$ are continuous but their derivatives diverge.
The transition point is called {\em singular sonic point}.
\item[Discontinuity] 
Both $c_s$ and $v_{\perp}$ are discontinuous as a consequence of discontinuity
in the equation of state. The point of discontinuity is called {\em shock}.
\end{description}

The second and the third case
are consequences of the idealization of
assuming an {\em irrotational inviscid perfect fluid}. 
 In reality, the divergences will be absent because one or more 
 simplifications become invalid as $v_{\perp}$ approaches $c_s$ or 
 at a shock. However, the mathematical divergences indicate that
 the corresponding divergent physical quantities, such as 
 analogue surface gravity, 
 may in a real physical situation be much larger than
 previously expected \cite{lib}.
 
\begin{figure}
\begin{center}
\includegraphics[width=.7\textwidth,trim= 0 0 0 0]{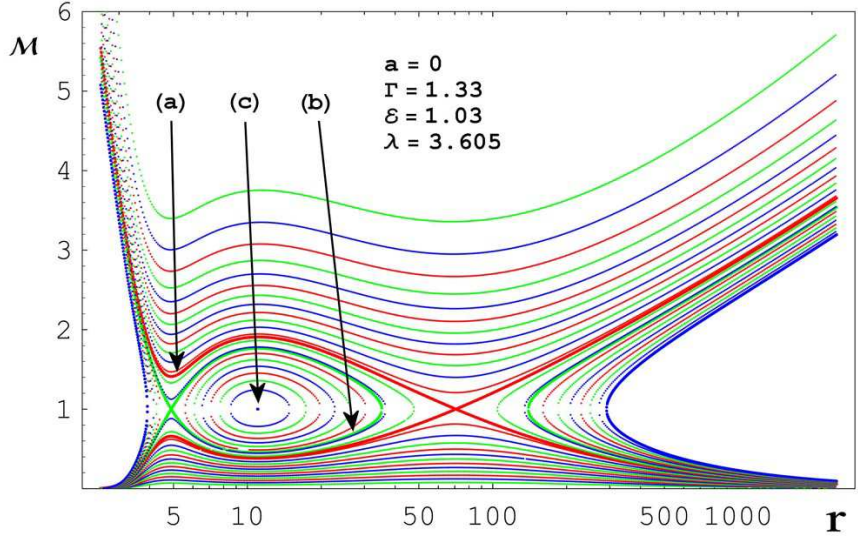}
\caption{
 Mach number $v_{\perp}/c_s$ versus $r$ (in units of the Schwarzschild radius)
 for various $\dot{N}/h$.
 % The sonic points
 %at which  
 %  the condition (\ref{eq215}) is satisfied are
 %indicated by the corresponding  $\dot{N}/h$. 
  The entropy measure  $\cal K$ corresponding to
 the flows marked by (a), (b) and (c) equals 
 0.0181289,
 0.0188906.
  0.0213815,
  respectively.     
}
\label{fig1}
\end{center}
\end{figure}
%%%%%%%%%%%%%%%%%%%%%%%%%%%%%%%%%
 
To find solutions with at least one regular sonic point, we
take the derivative  of (\ref{eq210})
and (\ref{eq211})  with respect to $r$, which yields
\begin{equation}
n'\frac{\dot{N}}{n} +n \left(\frac{\dot{N}}{n}\right)'=0,
\label{eq301}
\end{equation}
\begin{equation}
w'u_0 +w(u_0)'=0.
\label{eq302}
\end{equation}
%eliminating $n'/n$
%we may find another equationcondition
%\begin{equation}
%c_s^2\frac{(\dot{N}/n)'}{\dot{N}/n}-\frac{u_0)'}{u_0}=0
%\label{eq214}
%\end{equation}
Substituting $\dot{N}/n$ and $u_0$ from
(\ref{eq210})
and (\ref{eq211}) and
using (\ref{eq213}),
equations (\ref{eq301}) and (\ref{eq302}) may be written as
  a system of linear algebraic equations for
$w'/w$ and $v_{\perp}'/v_{\perp}$
\begin{equation}
\frac{w'}{w} +c_s^2\gamma_{\perp}^2\frac{v_{\perp}'}{v_{\perp}}
=-c_s^2\frac{1}{2}\frac{\Delta'}{\Delta}\, ,
\label{eq304}
\end{equation}
\begin{equation}
\frac{w'}{w} +v_{\perp}^2\gamma_{\perp}^2\frac{v_{\perp}'}{v_{\perp}}
=-\frac{1}{2}\frac{\Delta'}{\Delta}+\frac{1}{2}\frac{B'}{B}\, .
\label{eq305}
\end{equation}
Obviously, the determinant of the system vanishes at the acoustic
horizon
 owing to (\ref{eq208}). The system will have 
 a regular solution at the horizon if and only if
the right-hand sides of the two equations become equal
at the horizon.
In this case, the algebraic system (\ref{eq304})-(\ref{eq305})
degenerates at the horizon into one equation.
Hence, the condition
\begin{equation}
(1-c_s^2)\frac{\Delta'}{\Delta}-\frac{B'}{B}=0
\label{eq215}
\end{equation}
must hold at the horizon. 
If this condition is not satisfied, any solution to
 (\ref{eq304})-(\ref{eq305})
will have   no acoustic horizon at all
 or will have {\em irregular} acoustic horizons, i.e., such that the derivatives 
 $v_{\perp}'$ and $c_s'$ approach  infinity as one approaches the horizon.
 
 The regularity condition (\ref{eq215})
gives $c_s$ at the horizon in terms of
$r_{\rm h}$ but the actual value of $r_{\rm h}$ 
is yet unknown. To find $r_{\rm h}$, we need to specify the 
equation of state.
 Given the equation of
state,
we can express $w$ in terms of $c_s$,
 substitute $c_s$ from
(\ref{eq215}) into (\ref{eq211}) and  thus obtain an equation
for $r$.
For example, equation (\ref{eq217}), which follows from
 the polytropic equation of state (\ref{eq216}),  yields
\begin{equation}
{\cal E} \left(\Gamma-2 +\frac{\Delta\, B'}{B \Delta'}\right)=
(\Gamma-1)\left(\frac{\Delta'}{B'}\right)^{1/2}.
\label{eq218}
\end{equation}
A solution
to  this equation
for fixed $\lambda$ and $\cal E$
determines the  horizon radius $r_{\rm h}$.
Equation (\ref{eq218}) may, in general, have more than one solution.
  However, not all real solutions are physically acceptable.
 The positivity of $u_0^2$ in (\ref{eq003}) requires
\begin{equation}
g_{\phi\phi}+2\lambda g_{0\phi}+ \lambda^2 g_{00}\leq 0.
\end{equation}
For a given $\lambda$, this inequality fixes the range of
acceptable $r$.

%%%CHANGES IN THREE CAPTIONS BELOW 3/8/04

\begin{figure}
\begin{center}
\includegraphics[width=.7\textwidth,trim= 0 0 0 0]{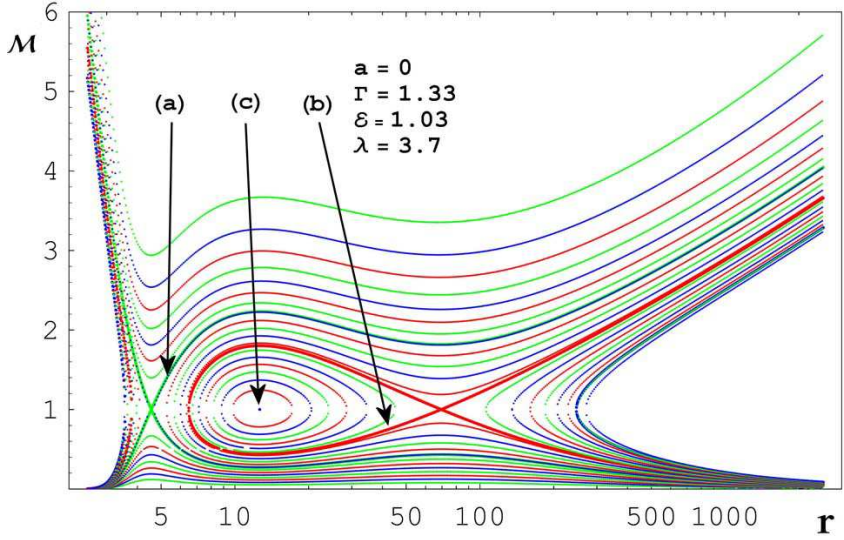}
\caption{
 Same as figure  \ref{fig1} for a different flow topology. 
 The entropy measure  $\cal K$ corresponding to
    (a), (b) and (c) equals 
 0.0158061,
 0.0180944
and 0.0206812,
respectively.        
}
\label{fig2}
\end{center}
\end{figure}
%%%%%%%%%%%%%%%%%%%%%%%%%%%%%%%%%

Once we have $r_{\rm h}$ and $c_s$ at the horizon, the
constant $\dot{N}/h$ may also be fixed from equation
(\ref{eq210}) evaluated at the horizon.
Equations (\ref{eq210}) and (\ref{eq211}) then
determine the
radial dependence
of $c_s$ and $v_{\perp}$.
It is important to emphasize that
this procedure determines only a subset
of solutions
 to equations  
(\ref{eq210}) and (\ref{eq211}) 
(or equivalently (\ref{eq304}) and (\ref{eq305}))
 which possess sonic points that
 satisfy the condition (\ref{eq215}).
 Any other solution will have 
  singular sonic points or
   no sonic point at all.
 This may be seen in figures  \ref{fig1} and \ref{fig2}
where two 
 typical sets of solutions
 with different topology
depending critically on $\lambda$
are plotted 
 for a fixed $\cal{E}$.
  Solutions are expressed in terms of the so-called Mach 
number defined as the ratio ${\cal M}\equiv v_{\perp}/c_s$. 
 The curves crossing the ${\cal M}=1$ line are transonic
and the crossing points are the location of the acoustic horizon.
The plots also include  
the configurations that 
do not satisfy (\ref{eq215}) at the acoustic horizon.
The plots describe both
wind and accretion, depending on the sign of $\dot{N}$ or
equivalently of $v_{\perp}$. Each line on the plot represents wind if
$\dot{N} <0$ ($v_{\perp}>0$) or accretion if
$\dot{N} >0$ ($v_{\perp}<0$).
The flow described by line (a) has an X-type sonic point.
The flow marked by (b) has two branches, each passing an X-type and 
ending at a singular sonic point.
An isolated O-type sonic point is marked by (c).

Obviously,
 a smooth flow will have at most one regular
  sonic point..
Hence, a smooth multitransonic flow is not possible.
However, a multitransonic flow with two sonic points
and a shock in between may be realized 
 if  discontinuous
transitions between different lines
are allowed. 

\begin{figure}
\begin{center}
\includegraphics[width=.7\textwidth,trim= 0 0 0 2cm]{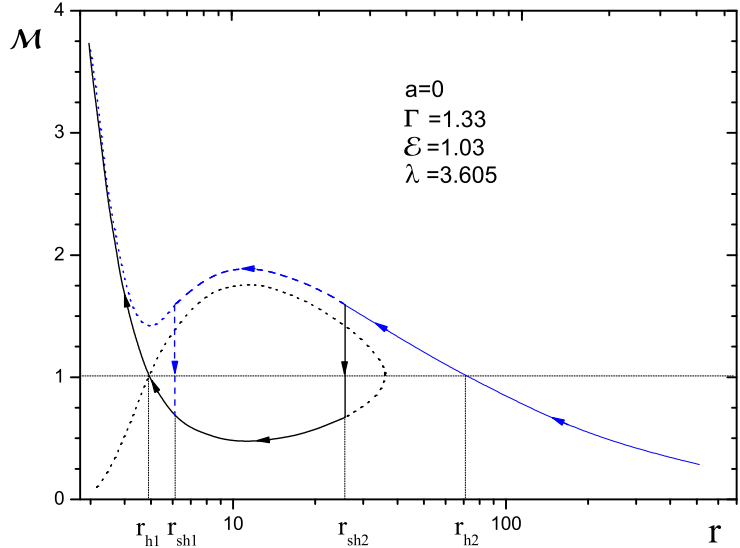}
\caption{
Transonic accretion with two sonic points
at $r_{\rm h1}$ and $r_{\rm h2}$
  and a shock in between  
at $r_{\rm sh2}$ (solid line).
The dashed line represents an alternative  flow with
a (unstable) shock at $r_{\rm sh1}$. The parameters are as in figure \ref{fig1}. 
}
\label{fig3}
\end{center}
\end{figure}

The lines in the plots in figures  \ref{fig1} and \ref{fig2}
in general correspond to different constants $\dot{N}/h$ 
and if the entropy is conserved, there is no way to jump from
one line to another. However, instead of keeping the constant
 ${\cal K}$  fixed, we  fix $\dot{N}/h$  and allow ${\cal K}$
 to vary from line to line. In this  way  the jumps between the lines 
 labelled by different ${\cal K}$ will be allowed at those points where
 the Rankine-Hugoniot conditions are satisfied.

The flows with singular
sonic points, i.e., the points where $v_{\perp}'$ and $c_s'$ diverge, will
formally have acoustic horizons with infinite surface gravity.
This property is generic for
inviscid flows \cite{lib} and the problem will
not appear in a realistic fluid with viscosity.
In our case, owing to the  shock formation, the  flow will actually never reach these points. 
We  assume that a stationary accretion flow goes from infinity
where it is subsonic to
the first unstable orbit of the black hole.
In the example depicted in figure \ref{fig1} 
the flow
 represented 
by line (a) 
passes through a regular sonic point.
If the Rankine-Hugoniot conditions are met, 
a jump from a supersonic  to a subsonic regime will take place, provided 
the entropy of the subsonic flow is larger.
Since ${\cal K}_{\rm (a)}< {\cal K}_{\rm (b)} < {\cal K}_{\rm (c)}$,
the entropy per particle $\sigma$,
 which increases with ${\cal K}$ according to (\ref{eq322}),
increases  going from line (a) 
towards the point denoted by (c) and
hence, a jump   
from (a) to (b) or to any of the closed loops
inside loop (b) is possible.
However, the only transition that will 
eventually result in a stable stationary flow
 will be a jump from (a) to the
subsonic branch of (b). A flow with a similar transition from (a) 
to one of the closed loops will inevitably 
end up at a singular sonic point, and hence cannot remain stationary. 

\begin{figure}
\begin{center}
\includegraphics[width=.7\textwidth,trim= 0 0 0 2cm]{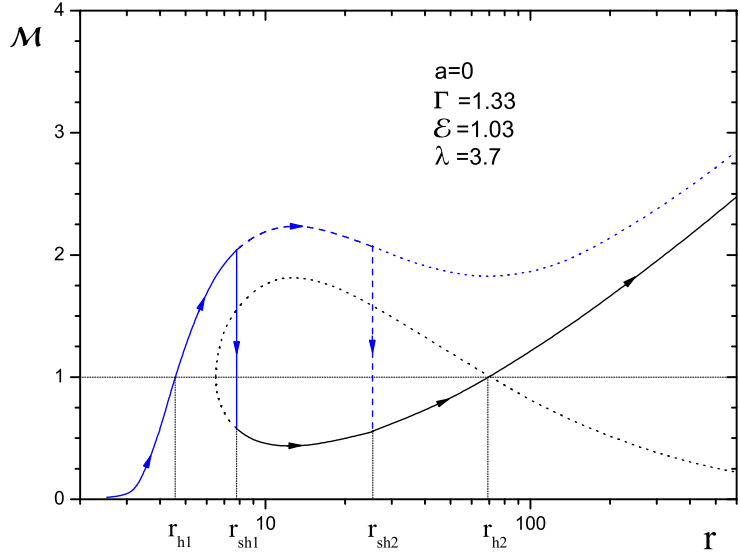}
\caption{
Transonic wind with two sonic points
at $r_{\rm h1}$ and $r_{\rm h2}$
  and a shock in between  
at $r_{\rm sh1}$ (solid line).
The dashed line represents an alternative flow with
a shock at $r_{\rm sh2}$. The parameters are as in figure \ref{fig2}. 
}
\label{fig4}
\end{center}
\end{figure}

The exact location of the shock $r_{\rm sh}$ for a particular flow configuration
may be calculated using the Rankine-Hugoniot conditions 
(\ref{eq401})-(\ref{eq403}) 
assuming that the outer flow is described by (a)
and the inner flow by the subsonic branch of (b). 
The first two conditions (\ref{eq401}) and (\ref{eq402}) 
  are trivially satisfied owing to
equations  (\ref{eq210}) and (\ref{eq211}). 
The particle number conservation 
(\ref{eq210}) across the shock may be written as
\begin{equation}
{\cal K}_1^{-\frac{1}{\Gamma-1}} \dot{\cal N}_1=
{\cal K}_2^{-\frac{1}{\Gamma-1}} \dot{\cal N}_2 \, ,
\label{eq404}
\end{equation}
where the subscripts 1 and 2 denote, respectively, the inner and outer
boundary values at the shock and
\begin{equation}
\dot{\cal N} \equiv
\eta  \left(\frac{\Gamma-1}{c_s^2}-1\right)^{-\frac{1}{\Gamma-1}}
v_{\perp}\gamma_{\perp}
\label{eq405}
\end{equation}
is a  kinematic quantity the change of which across the shock
depends on the change  of
 $v_{\perp}$ and $c_s$ only. Hence, to make the particle number
 conserved, a discontinuity  of $\dot{\cal N}$ is compensated by
 a corresponding discontinuity of ${\cal K}$ such that 
 equation (\ref{eq404}) holds.
 As a consequence, the entropy per particle $\sigma$
 being a continuous function of ${\cal K}$
 is not conserved
 across the shock.
 The third Rankine-Hugoniot condition
 (\ref{eq403}) may now be written as
\begin{eqnarray}
 & &
\eta^{\Gamma+1}
\left(\frac{\Gamma-1}{c_{s1}^2}-1\right)^{-\frac{1}{\Gamma-1}}
+(\Gamma-1) c_{s1}^2
\left(\frac{\Gamma-1}{c_{s1}^2}-1\right)^{-\frac{\Gamma}{\Gamma-1}}
\dot{\cal N}_1^2=
 \nonumber \\
% & & \nonumber \\
%\!&\!=&    \nonumber \\
  & &
\left[\eta^{\Gamma+1}
\left(\frac{\Gamma-1}{c_{s1}^2}-1\right)^{-\frac{1}{\Gamma-1}}
+(\Gamma-1) c_{s2}^2
\left(\frac{\Gamma-1}{c_{s2}^2}-1\right)^{-\frac{\Gamma}{\Gamma-1}}
\dot{\cal N}_2^2\right] \left(
\frac{{\cal K}_1}{{\cal K}_2}\right)^{\frac{1}{\Gamma-1}} ,
% & &
\label{eq406}
\end{eqnarray}
and may be solved for $r$. 
In figures  \ref{fig3} and \ref{fig4} we show the solutions for the  
accretion and wind, respectively, corresponding to the flows depicted in figures  \ref{fig1} and  \ref{fig2}.
The two possible locations of the shock are shown by
vertical lines at $r_{\rm sh1}$ and $r_{\rm sh2}$
with arrows indicating the flow orientation.
According to a standard
local stability analysis \cite{yan},
one can show for a multitransonic accretion that
only the shock formed in  between 
the middle (indicated by (c) in 
figures \ref{fig1} and \ref{fig2})
 and the outer sonic point is stable.
 Hence, in our accretion example in figure \ref{fig3},
 the shock at $r_{\rm sh2}$ is stable and the one
 at  $r_{\rm sh1}$ is unstable.

\section{Acoustic metric and causal structure}
\label{causal}
\begin{figure}
\begin{center}
\includegraphics[width=.95\textwidth,trim= 0 0 0 0]{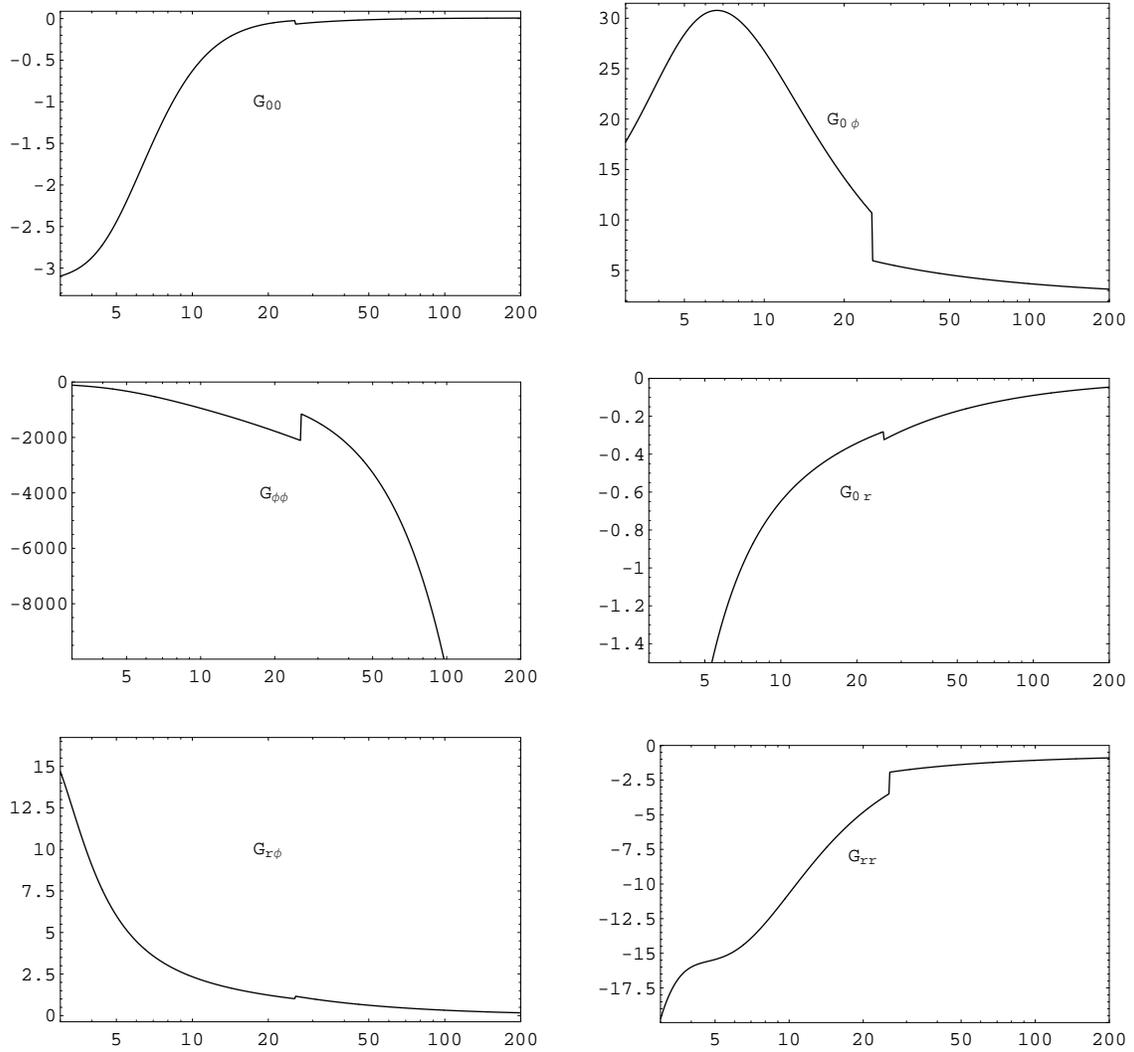}
\caption{
 The components of the acoustic metric for the accretion flow
 as in figure \ref{fig3}
 }
\label{fig00}
\end{center}
\end{figure}

\begin{figure}
\begin{center}
\includegraphics[width=.95\textwidth,trim= 0 0 0 0]{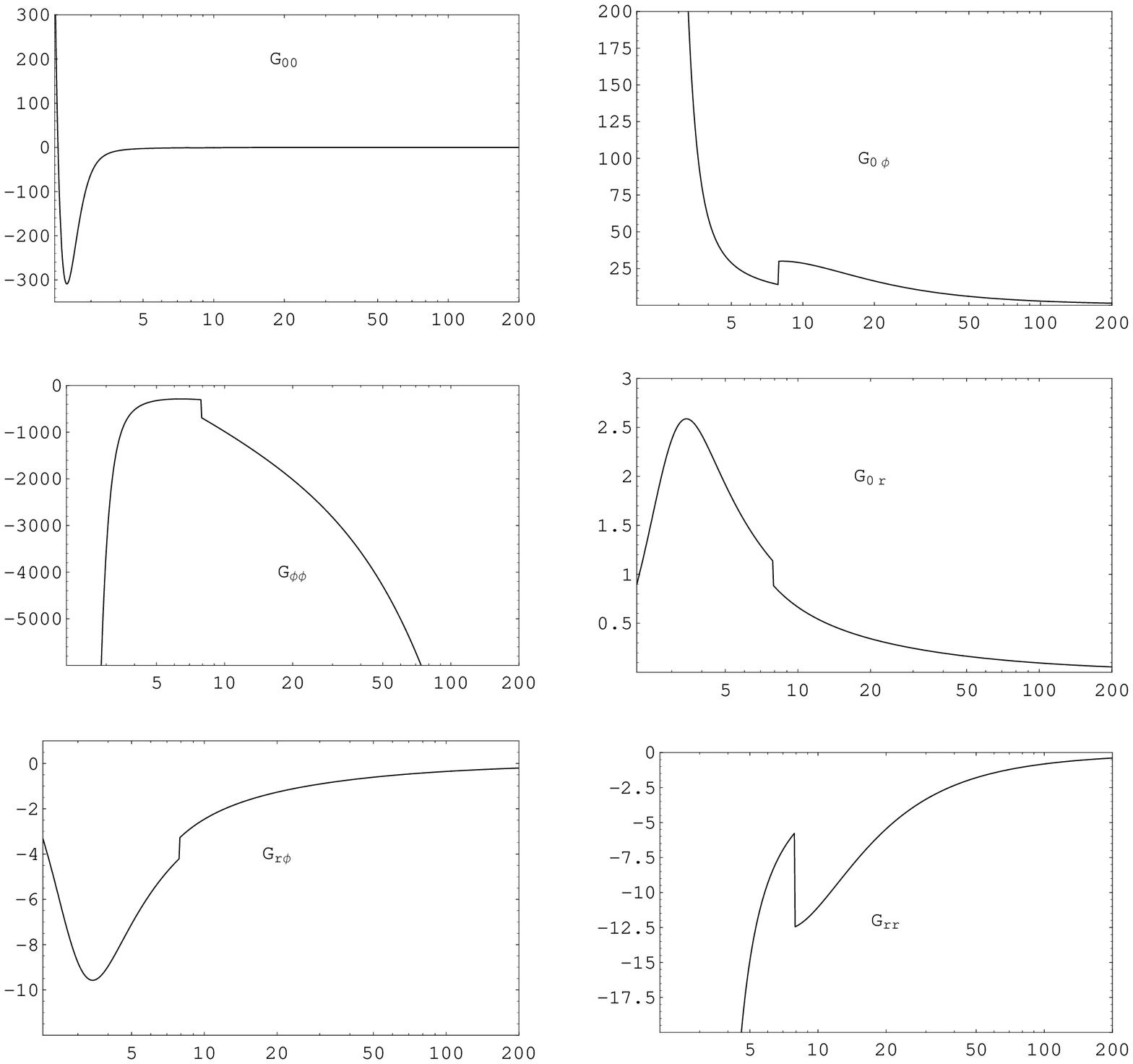}
\caption{
 The components of the acoustic metric for the accretion flow
 as in figure \ref{fig4}
 }
\label{fig00b}
\end{center}
\end{figure}

\begin{figure}
\begin{center}
\includegraphics[width=.7\textwidth,trim= 0 0 0 2cm]{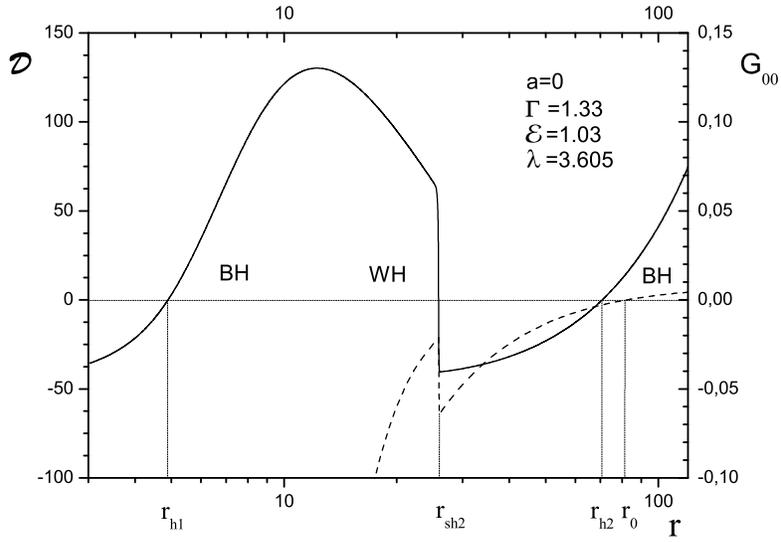}
\caption{
Acoustic metric functions $\cal D$ (solid line) and $G_{00}$ 
(dashed line) 
versus $r$
for the multitransonic accretion as in figure \ref{fig3}
with a  shock   
at $r_{\rm sh2}$. 
 Acoustic horizons
 and their type (from the point of view of an observer
 in the subsonic region)
  are indicated by BH (black hole) and 
 WH (white hole).  
}
\label{fig5}
\end{center}
\end{figure}

\begin{figure}
\begin{center}
\includegraphics[width=.7\textwidth,trim= 0 0 0 2cm]{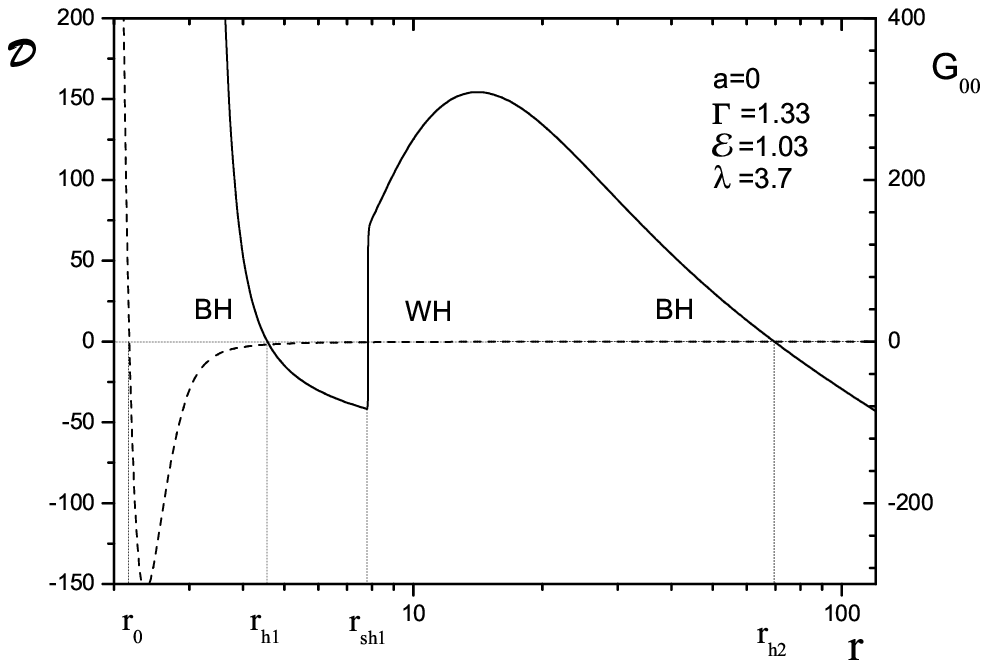}
\caption{
Acoustic metric functions $\cal D$ (solid line) and $G_{00}$ 
(dashed line) 
versus $r$
for the multitransonic wind  as in figure \ref{fig4}
with a  shock   
at $r_{\rm sh1}$. 
}
\label{fig6}
\end{center}
\end{figure}

The geometric aspects of the axially symmetric accretion
are conveniently described in terms of
the acoustic metric defined by (\ref{eq029}) and
the metric discriminant ${\cal D}$ defined by
(\ref{eq209}). 
In figure \ref{fig00}
 we plot the components of the metric tensor
as functions of $r$
calculated using equation (\ref{eq029}) for the transonic accretion flow represented by the solid line in figure \ref{fig3} and 
similarly in \ref{fig00b} for the transonic wind of figure
\ref{fig4}.
For the same 
flow configurations depicted in figures  \ref{fig3} and
\ref{fig4} we also plot
${\cal D}$ as a function of $r$
in  figures  \ref{fig5} and \ref{fig6}, 
 respectively.
Clearly, ${\cal D}$ is positive (negative) in the subsonic
(supersonic) region.
Acoustic horizons are located at the points where
the discriminant changes its sign.
Depending on the slope of  ${\cal D}$ at these points 
passing in the direction of the flow,
we call them acoustic {\em black hole} 
(if the slope is descending)
or {\em white hole} 
(if the slope is ascending)
 horizons. 
Indeed, for both the multitransonic accretion
(figure \ref{fig3}) and the wind (figure \ref{fig4}), 
the flow at the inner and the outer sonic points
denoted by $r_{\rm h1}$ and $r_{\rm h2}$, respectively,
goes from subsonic to supersonic; so,
according to the classification of Barcel\'o et al
\cite{barc2},
there we have acoustic black hole horizons.
 At the discontinuity  $r_{\rm sh1}$ or $r_{\rm sh2}$
    the flow goes from supersonic to subsonic
 and hence observers in the subsonic region of the accretion
 (between $r_{\rm h1}$ and $r_{\rm sh2}$ in figures 
 \ref{fig3} and \ref{fig5} ), and of the wind
 (between
$r_{\rm sh1}$ and $r_{\rm h2}$ in figures 
 \ref{fig4} and \ref{fig6}) experience a white hole. 

%%%%%%%%%%%%%%%%%%%%%%%%%%%%%%%%% 
%%%%%%%%%%%%%%%%%%%%%%%%%%%%%%%%% 


 In  figures  \ref{fig5} and \ref{fig6} we also plot 
 the metric component $G_{00}$ as a function of $r$ 
 (represented by the dashed line).
 The point $r_0$ where $G_{00}$ changes its sign 
is the radius of the stationary limit surface and the boundary of the acoustic ergo region. 

\begin{figure}
\begin{center}
\includegraphics[width=.7\textwidth,trim= 0 0 0 0]{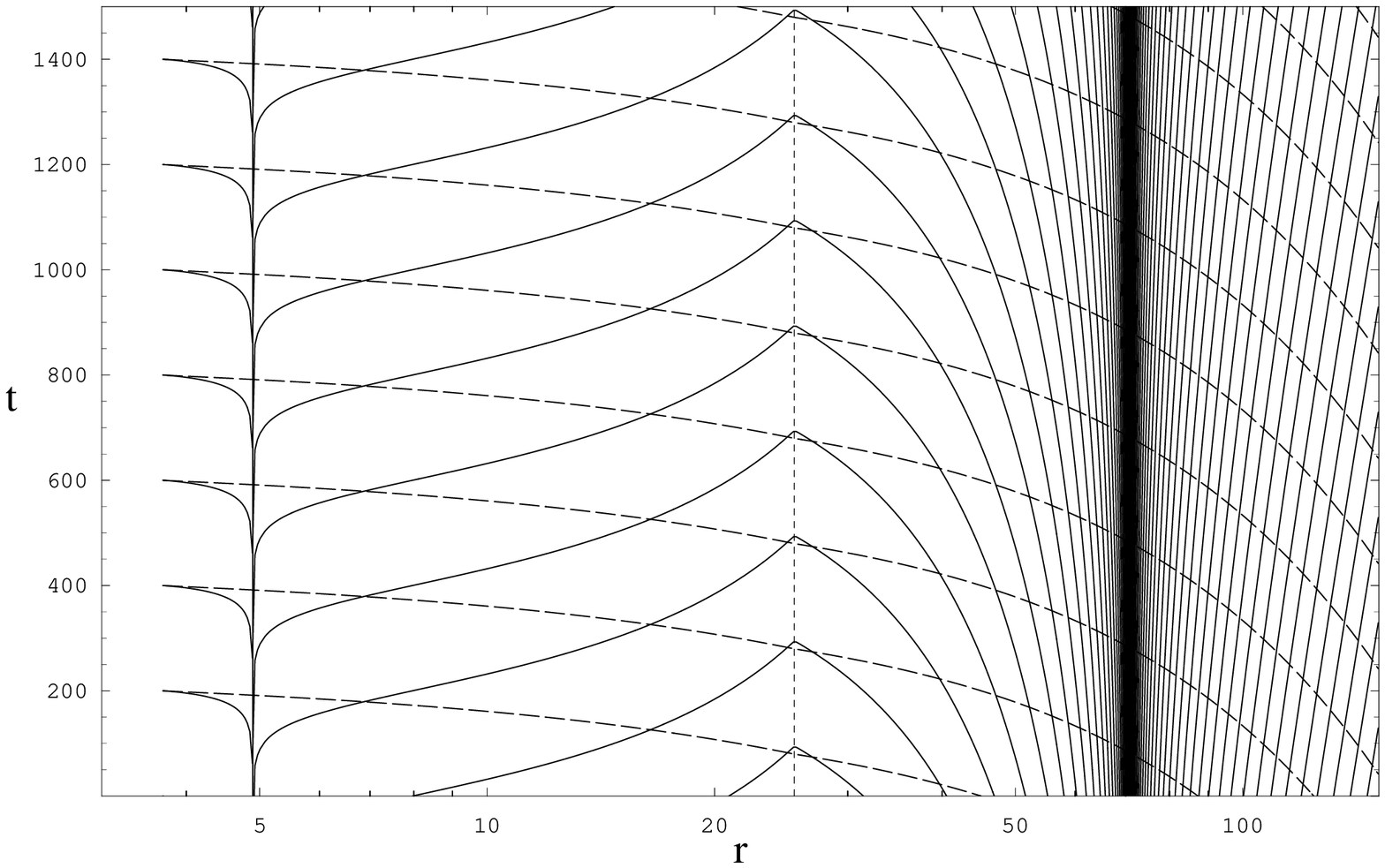}
\caption{
 Acoustic causal structure of the transonic accretion of figure 
 \ref{fig3} with the null curves $u={\rm const}$ (solid lines) 
 and $w={\rm const}$ (dashed lines).
}
\label{fig0}
\end{center}
\end{figure}
%%%%%%%%%%%%%%%%%%%%%%%%%%%%%%%%%
\begin{figure}
\begin{center}
\includegraphics[width=.7\textwidth,trim= 0 0 0 0]{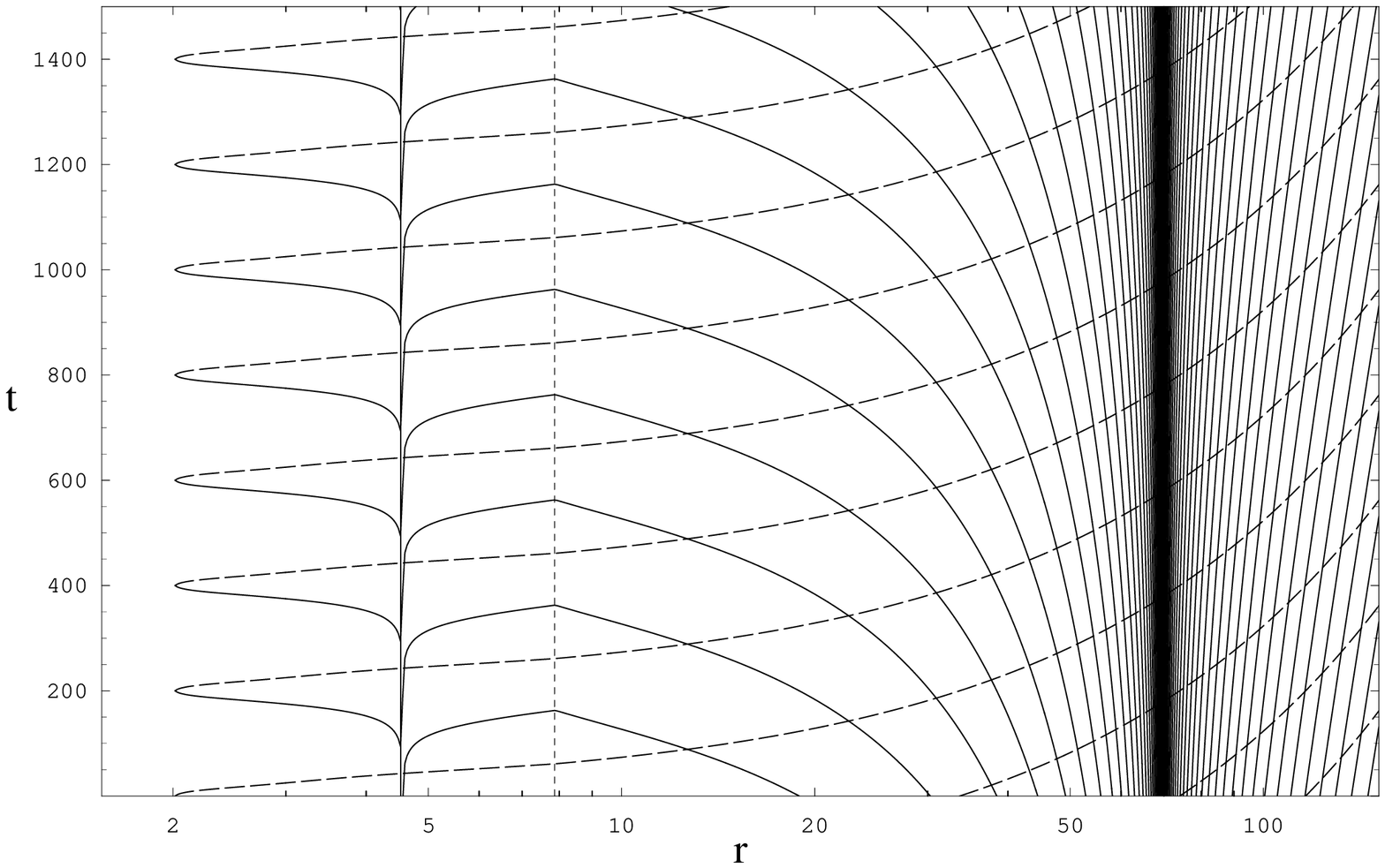}
\caption{
 Acoustic causal structure of the transonic wind of figure 
 \ref{fig4} with the null curves $w={\rm const}$ (solid lines) 
 and $u={\rm const}$ (dashed lines).
}
\label{fig0b}
\end{center}
\end{figure}
%%%%%%%%%%%%%%%%%%%%%%%%%%%%%%%%%

In order to study the causal structure of the 
axially symmetric acoustic geometry, let us introduce 
retarded and advanced radial null coordinates $u$ and $w$, respectively.
We first make a transformation to the acoustic corotating frame
by eliminating the term $G_{0\phi}dtd\phi$ from the acoustic metric.
This is achieved by  transforming $\phi$ to $\phi'$ as
\begin{equation}
d\phi'=d\phi+\frac{G_{0\phi}}{G_{\phi\phi}}dt \, ,
\label{eq047}
\end{equation}
with the remaining coordinates unchanged. This yields
 the acoustic metric
 \begin{equation}
ds^2=G_{\mu\nu}'dx^{\mu}dx^{\nu} ,
\label{eq048}
\end{equation} 
where 
\begin{equation}
G_{00}'=-\frac{\cal D}{G_{\phi\phi}};
\;\;\;\;\;\;
G_{0\phi}'=0,
\label{eq049}
\end{equation} 
\begin{equation}
G_{0r}'=G_{r0}'=G_{0r}-G_{r\phi}\frac{G_{0\phi}}{G_{\phi\phi}},
\label{eq050}
\end{equation} 
with all remaining components of the acoustic metric  unchanged.
We then make the transformation 
\begin{equation}
du=dt+ \frac{1}{c_+}dr \, ,
\label{eq051}
\end{equation} 
\begin{equation}
dw=dt+ \frac{1}{c_-}dr \, ,
\label{eq052}
\end{equation} 
where
\begin{equation}
c_{\pm}=\frac{G_{0r}'\pm \sqrt{G_{0r}'^2-G_{00}'G_{rr}}}{G_{rr}} \, .
\label{eq053}
\end{equation}
The acoustic metric at constant $\phi'$ is then given by
\begin{equation}
\left. ds^2\right|_{\phi'={\rm const}}=-\frac{\cal D}{G_{\phi\phi}}du dw \, .
\label{eq054}
\end{equation} 
The functions $t=t(r)$
at constant $u$
or constant $w$, 
obtained by integrating
\begin{equation}
t=- \int \frac{1}{c_{\pm}}dr \, ,
\label{eq055}
\end{equation} 
 represent, respectively, the right-moving and left-moving sound
rays or {\em null-rays} \cite{barc2}.

In figures \ref{fig0} and \ref{fig0b} we plot
the  $u={\rm const}$
and $w={\rm const}$  lines for the accretion  of 
figure \ref{fig3} and the wind of figure \ref{fig4}.
The divergence of lines at the acoustic horizons
towards $t=-\infty$ demonstrate an acoustic black hole.
Since the transition from 
a supersonic to a subsonic region is discontinuous at the shock,
the lines have a cusp pointing in the direction of positive $t$.
If the transition at the shock were smooth,
the null lines would diverge towards $t=+\infty$ as in the 
white-hole example of 
 Barcel\'o et al
\cite{barc2}.
Hence, the cusp indicates a white hole.

%%%%%%%%%%%%%%%%%%%%%%%%%%%%%%%%%

\section{Surface gravity and Hawking temperature} 
\label{surface}
One of the most interesting aspects of acoustic horizons is
the existence of a surface gravity and the associated Hawking radiation. 
To calculate the analogue surface gravity, we  need
the derivatives $c_s'$ and $v_{\perp}'$  at the
acoustic horizon.
Let us consider a regular sonic point, e.g.,
at $r_{\rm h1}$ or $r_{\rm h2}$
in figures \ref{fig3}-\ref{fig6}, first.
The case of shock will be discussed afterwards.
To calculate $c_s'$ and $v_{\perp}'$ at a regular sonic point,
we have to solve two equations.
The first one is obtained combining  (\ref{eq304}) and the derivative of (\ref{eq217}):
\begin{equation}
\frac{4c_s'}{\Gamma -1-c_s^2}+
\frac{2 v_{\perp}'}{1-c_s^2}=
-c_s\frac{\Delta'}{\Delta}\, .
\label{eq214}
\end{equation}
The second equation is obtained by
eliminating $w'/w$ from (\ref{eq304}) and (\ref{eq305}) and
taking the derivative
with respect to $r$
of the equation thus obtained at the horizon.
%\begin{equation}
%\frac{v_{\perp}'}{v_{\perp}}\gamma_{\perp}^2(c_s^2-v_{\perp}^2)=
%\frac{1}{2}(1-c_s^2)\frac{\Delta'}{\Delta}-\frac{1}{2}\frac{B'}{B}
%\label{eq219}
%\end{equation}
 We find
\begin{equation}
4 v'_{\perp}\gamma^{2}_{\perp}(c'_{s}-v'_{\perp})
+2 c_s c'_s\frac{\Delta'}{\Delta} =
\left(1-c^{2}_{s}\right)
\left(\frac{\Delta''}{\Delta}-\frac{\Delta'^2}{\Delta^2}\right)
-\frac{B''}{B}+\frac{B'^2}{B^2}\, .
\label{eq220}
\end{equation}
Equations  (\ref{eq214}) and (\ref{eq220}) can now be easily solved
for $c_s'$ and $v_{\perp}'$.

The analogue surface gravity $\kappa$
may  be calculated with help of
the Killing
field $\chi^{\mu}$ that is null on the horizon.
We start from the expression \cite{bil,wal}
\begin{equation}
 G^{\mu\nu}
 n_{\nu}\frac{\partial}{\partial n}
( G_{\alpha\beta} \chi^{\alpha}\chi^{\beta}) =
 2\kappa \chi^{\mu} ,
\label{eq141}
\end{equation}
where $\partial/\partial n\equiv n^{\mu}\partial_{\mu}$
 denotes the normal
derivative at the horizon.
Since the definition
of surface gravity by
equation (\ref{eq141}) is conformally invariant
\cite{jac93},
in the calculations that follow
 we drop the conformal factor
$n/(w c_s)$ in
 $G_{\mu\nu}$.
 
 %%%%%%%%%%%%%%%%%%%%%%%%%%%%%%%%%
\begin{figure}
\begin{center}
\includegraphics[width=.9\textwidth,trim= 0 0 0 2cm]{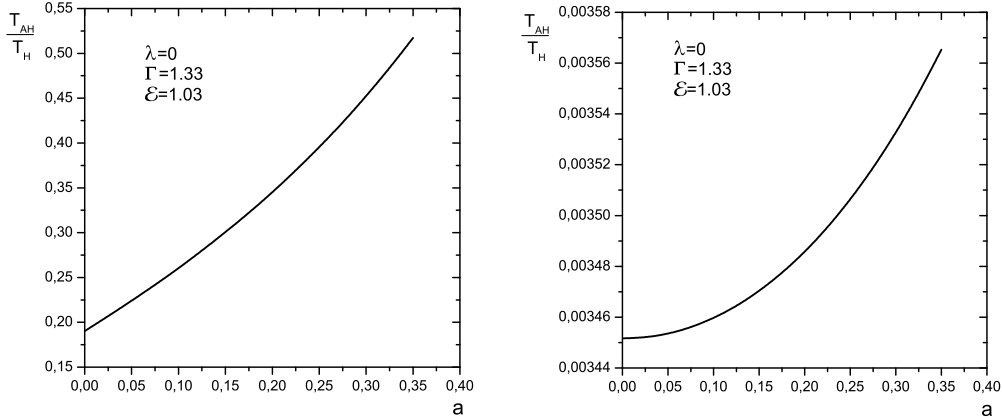}
\caption{
Analogue Hawking temperature at the inner (left)
and outer (right)
acoustic  horizon versus the specific angular momentum of
the black hole.
}
\label{fig7}
\end{center}
\end{figure}

%%%%%%%%%%%%%%%%%%%%%%%%%%%%%%%%%
Consider  the vector field
 $L^{\mu}$ defined by
(\ref{eq237}).
 Its magnitude in the acoustic metric is given by
\begin{equation}
\parallel\! L\! \parallel^2 \equiv G_{\mu\nu} L^\mu L^\nu =
 [ 1-(1-c_s^2)\gamma_{\perp}^2]=
\gamma_{\perp}^2 (c_s^2-v_{\perp}^2) .
\label{eq046}
\end{equation}
Hence, the vector $L^{\mu}$ is null on the horizon.
We  now construct a Killing vector $\chi^{\mu}$
in the form
\begin{equation}
\chi^{\mu}=\xi^{\mu} +\Omega_{\rm h} \phi^{\mu} ,
\label{eq147}
\end{equation}
where the Killing vectors
$\xi^\mu$ and $\phi^{\mu}$
are the generators of the temporal and axial isometry group, respectively.
It may be easily verified that if the constant $\Omega_{\rm h}$
coincides with the 
angular velocity 
(\ref{eq001}) evaluated at the acoustic horizon, then
$\chi^{\mu}$ becomes
parallel to $L^{\mu}$
and hence null
on the horizon.
Using
 $\chi^{\mu}=
\sqrt{\chi^{\nu}
\chi_{\nu}}
L^{\mu}$ and the definition (\ref{eq237})
we find
\begin{equation}
 G^{\mu\nu} n_{\nu}   =
 n^{\mu}-\frac{1}{v_{\perp}\gamma_{\perp}}u^{\mu}=
 -\frac{1}{v_{\perp}}
  \frac{\chi^{\mu}}{\sqrt{\chi^{\nu}\chi_{\nu}}}
\label{eq153}
\end{equation}
at the horizon. 
Equation (\ref{eq141}), together with (\ref{eq046}) and (\ref{eq153})
yields
 \begin{equation}
\kappa=\frac{1
  }{1-c_s^2}
\sqrt{\frac{\chi^{\nu}\chi_{\nu}}{-g_{rr}}}(v_{\perp}'-c_s'),
\label{eq043}
\end{equation}
where it is understood that
 the derivatives and
 the quantity
 $\chi^{\mu}\chi_{\mu}$ are to be taken
 at the horizon.
The corresponding Hawking temperature
$T_{\rm AH}=|\kappa|/(2\pi)$ represents
the temperature as measured by an observer near
 infinity.
For $\lambda=a=0$, equation (\ref{eq043}) reduces to
(\ref{eq238}) derived in a different way for
a nonrotating fluid in a static background metric.

The norm of the Killing vector $\chi_{\mu}$ is 
easily calculated from (\ref{eq147}) using (\ref{eq002}).
We find
\begin{equation}
\sqrt{\chi^{\mu}\chi_{\mu}}=
\frac{r\sqrt{\Delta \, B}}{r^{3}+a^{2}r+2Ma^{2}
-2\lambda Ma}.
\label{eq307}
\end{equation}

\begin{figure}
\begin{center}
\includegraphics[width=.9\textwidth,trim= 0 0 0 2cm]{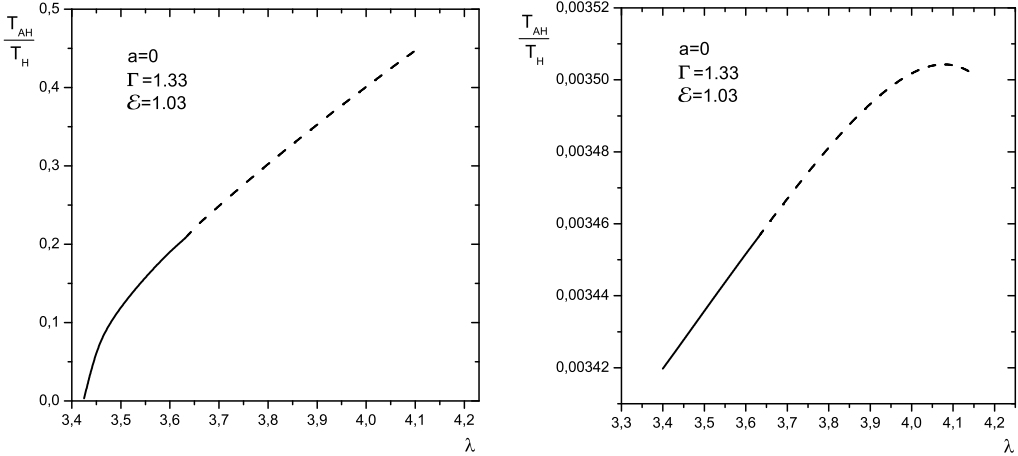}
\caption{
Analogue Hawking temperature at the inner (left)
and outer (right)
acoustic  horizon versus the specific angular momentum of
the disk for the two 
topologies depicted in figure \ref{fig1} (solid line)
and figure \ref{fig2} (dashed line).
}
\label{fig8}
\end{center}
\end{figure}

We plot the ratio of the analogue 
Hawking temperature $T_{\rm AH}$ to 
the black hole Hawking temperature $T_{\rm H}$
at the inner and outer horizons as a function 
of the  black-hole specific angular momentum $a$
(figure  \ref{fig7})
and  of  
the disk specific angular momentum $\lambda$
(figure  \ref{fig8}).
The solid and dashed lines represent the two topologies of
figures \ref{fig1} and \ref{fig2}, respectively.
However, whereas  
the inner and the outer acoustic  horizons 
for the topology of figure \ref{fig1} belong
to the same multitransonic flow connected by a shock
(the typical configuration depicted in figure
\ref{fig3}),
the horizons for
the topology of figure \ref{fig2} 
belong to disconnected transonic flows
 marked by  (a) and (b) in figure \ref{fig2}.
 Amazingly, the analogue surface gravity
is smooth  at the 
 topological transition point.
 
Note that the
  analogue Hawking temperature $T_{\rm AH}$ of the outer horizon
  is much lower than
  the temperature $T_{\rm AH}$ of the inner horizon, which in turn is,
      although close in
magnitude, typically lower than 
the black hole Hawking temperature $T_{\rm H}$.
This is due to the relatively low specific energy ${\cal E} =1.01$,
barely exceeding the rest mass.
For a range
of larger values of $\cal E$ , e.g., close to 2, the temperature 
$T_{\rm AH}$ would exceed $T_{\rm H}$ as in the 
spherically symmetric accretion \cite{dascqg,dasgrg}.
However, we have found that in this range of large specific energies
a multitransonic behaviour is absent.

%%%%%%%%%%%%%%%%%%%%%%%%%%%%%%%%%
 Calculation of the surface gravity at the shock
 is not possible in the present model with no viscosity. 
The shock may be viewed as
a boundary separating two phases of the fluid.
The equation of state, and hence the sound speed,
 as well as 
 the radial component of the fluid velocity,
 in each phase
are different and exhibit discontinuity at the boundary.
As a consequence, 
the surface gravity and the associated Hawking temperature at
 the shock will be formally infinite.
 However, in a more realistic
situation, the phase  boundary, owing to viscosity, will have a finite
thickness and discontinuities will be smoothed out.
The gradients of $v_{\perp}$ and $c_s$ will actually be finite but
possibly very large.
A similar situation occurs in a non-relativistic
acoustic setup in which the boundary of a stable-phase bubble
propagates in the surrounding metastable phase
\cite{vac} and 
in the $^3$He-A superfluid
where a superluminally moving domain wall soliton provides
an analogue event horizon
\cite{jac1,jac2}.
 
 \section{Acoustic superradiance}
 \label{superradiance}
The existence of an ergo region 
in the transonic accretion flow 
 as described in sections \ref{acoustic} and \ref{causal} 
suggests the possibility
of {\em acoustic superradiance}.
 Acoustic waves entering the ergo region and
reflecting from the outer acoustic horizon may experience a coefficient 
of reflection greater than 1. This effect 
is a close analogue of superradiant scattering \cite{zel,sta}
which allows energy extraction from  Kerr black holes.
Acoustic superradiance has recently been extensively discussed in
the context of the draining bathtub \cite{bas2,ber} and its application to
the Bose-Einstein condensate \cite{bas}.
 
The quantitative investigation of acoustic superradiance 
in the accretion disc geometry may be performed in the usual way:
First, the variables  in (\ref{eq028}) may be separated using a cylindrical 
 wave ansatz
\begin{equation}
\varphi(t,r,\phi)= \Phi(r) e^{-i \omega t} e^{i m \phi}\, ,
\label{eq155}
\end{equation} 
where $\omega$ is real and positive  and $m$ a positive integer.
Equation (\ref{eq028}) then reduces to an ordinary 
second-order differential equation for 
$\Phi(r)$ which may be solved numerically for suitable boundary
conditions. However, this calculation may be quite involved for 
the transonic flow  
described in sections \ref{acoustic}  and \ref{causal}
and would go
beyond the scope of this paper.
Instead,
we would  like to demonstrate here that 
the acoustic superradiance occurs
in a relativistic transonic axisymmetric flow 
 if 
\begin{equation}
0<\omega < m\Omega_{\rm h} \, ,
\label{eq156}
\end{equation} 
 where $\Omega_{\rm h}$ is the flow angular velocity (\ref{eq001})
evaluated at the acoustic horizon. To show this, we follow the
procedure outlined in \cite{wal}.
 
Using the stress tensor (\ref{eq329}) and the stationary Killing vector 
$\xi^{\mu}$
 we first construct the analogue ``energy current"
\begin{equation}
J_{\mu}=T_{\mu\nu} \xi^\nu; \;\;\;\;\;
J^\mu=G^{\mu\nu} J_{\nu}\, ,
\label{eq157}
\end{equation}
 which is conserved, i.e. which satisfies $\nabla_{\mu} J^\mu =0$,
$\nabla_{\mu}$ being the covariant derivative associated with $G_{\mu\nu}$.
 Hence if we integrate $\nabla_{\mu} J^\mu $ over the region $S$ with  
 the boundary
  $\partial S$, we find by Gauss's law
  \begin{equation}
\int_{\partial S} dS_{\mu} J^{\mu}=
\int_{C} dS_{\mu} J^{\mu} +
\int_{\Sigma_1} dS_{\mu} J^{\mu}+
\int_{\Sigma_2} dS_{\mu} J^{\mu}+
\int_{\Delta{\mathcal H}} dS_{\mu} J^{\mu}=0 ,
\label{eq158}
\end{equation}
where we have chosen $\partial S$ such that it consists of
  the cylinder $C$ 
 of large radius $R\rightarrow\infty$ and height $\Delta t$, the disc 
 $\Sigma_1$ at the bottom of the cylinder
 extended from the outer acoustic horizon radius $r_{\rm h}$  to $R$,
  the time translated disk 
  $\Sigma_2$ at the top,  
   and the 
  part $\Delta{\mathcal H}$ of the outer acoustic horizon
  ${\mathcal H}$ between $\Sigma_1$ and $\Sigma_2$.
  The integrals over $\Sigma_1$ and $\Sigma_2$ cancel by time symmetry,
  the integral over $C$ represents the net energy flow out of
  $S$ to infinity, i.e., the difference between the outgoing and incoming
  energies $E_{\rm out}-E_{\rm in}$ during the time $\Delta t$,
  whereas the integral over $\Delta{\mathcal H}$ represents the net energy
  flow into the acoustic black hole.
  Thus we have
   \begin{equation}
 E_{\rm out}-E_{\rm in} =  
 -\int_{\Delta{\mathcal H}} dS_{\mu} J^{\mu}
=\int dA dt \,\chi^{\mu} J_{\mu} \, ,
\label{eq159}
\end{equation} 
where we have used $n_{\mu}=-G_{\mu\nu} \chi^\nu$ as the outward directed
normal to the horizon and
$\chi^{\mu}$ is given by (\ref{eq147}).
 The integral in (\ref{eq159}) is taken over the proper volume
element $dAdt$ on the acoustic horizon.
Using $G_{\mu\nu}\chi^\mu\xi^\nu=0$ which follows directly from
(\ref{eq153}),
a straightforward algebra yields
\begin{equation}
 \chi^{\mu} J_{\mu}=
 \chi^{\mu}\partial_\mu\varphi \, \xi^{\nu}\partial_{\nu}\varphi = 
 (\partial_t \varphi +\Omega_{\rm h} \partial_\phi\varphi) \partial_t\varphi. 
\label{eq160}
\end{equation} 
 Applying this to (\ref{eq155}) it follows from (\ref{eq159}) 
 that the time averaged energy flux across the horizon is
 given by
 \begin{equation}
  \frac{1}{\Delta t}(E_{\rm in}-E_{\rm out})
 =\frac{1}{2}A\,\Phi^2 \omega(\omega-m\Omega_{\rm h})\, ,
 \label{eq161}
\end{equation} 
 where $A$ is the proper ``area" of the horizon.
Thus, in the frequency range of equation (\ref{eq156}),
the energy loss across the horizon is negative and hence
superradiance is obtained.

 \section{Conclusions}
 \label{conclude}
 In this paper we have studied the phenomena related
 to the acoustic geometry of the axially symmetric
 accretion flow. 
 We have discussed the formation of the shock and acoustic horizons 
 using a general relativistic formalism
 in a model of the flow based on a perfect
 polytropic gas.
 We have found that a multitransonic flow can form with
 two sonic points and a shock in between.
 Null curves depicted in figures 
 \ref{fig0} and \ref{fig0b} clearly demonstrate the
 presence of the acoustic black hole at regular sonic points
 and of the white hole at the shock.
 
 Multitransonic accretion happens for a range of values of the specific energy
 density not much exceeding the rest mass. 
 The analogue Hawking temperature of the inner horizon, calculated for
 a typical set of physical parameters, is lower but close in magnitude to
 the black-hole Hawking temperature. However, at the middle horizon that forms
 at the shock location,
  the surface gravity is formally infinite owing
 to discontinuities in the speed of sound and the radial velocity.
  We expect that in a realistic fluid with viscosity,  discontinuity will
  be removed and replaced by a steep slope resulting in a finite
  but possibly very high analogue Hawking temperature. 
    
\section*{Acknowledgments}
 The work of HA and NB   was supported by
 the Ministry of Science and Technology of the
 Republic of Croatia under Contract
 No.\ 0098002 and partially supported through the
 Agreement between the Astrophysical sector, SISSA, and the
 Particle Physics and Cosmology Group, RBI.

\end{document}